\documentclass[fleqn,usenatbib]{mnras}

\usepackage{newtxtext,newtxmath}
\usepackage{soul}

\usepackage[T1]{fontenc}

\DeclareRobustCommand{\VAN}[3]{#2}
\let\VANthebibliography\thebibliography
\def\thebibliography{\DeclareRobustCommand{\VAN}[3]{##3}\VANthebibliography}
\usepackage[dvipsnames]{xcolor}

\usepackage{graphicx}	
\usepackage{amsmath}	

\title[High-accuracy cosmological emulation]{
High-accuracy emulators for observables in $\Lambda$CDM, $N_\mathrm{eff}$, $\Sigma m_\nu$, and $w$ cosmologies
}

\author[]{Boris Bolliet,$^{1}$\thanks{E-mail: boris.bolliet@gmail.com} Alessio Spurio Mancini,$^{2}$\thanks{E-mail: a.spuriomancini@ucl.ac.uk} J.~Colin Hill,$^{3}$ Mathew Madhavacheril,$^{4}$ Hidde T. Jense,${^5}$ \newauthor Erminia Calabrese,${^5}$ Jo Dunkley${^{6,7}}$
 \\
$^{1}$DAMTP, Centre for Mathematical Sciences, Wilberforce Road, Cambridge CB3 0WA, UK\\
$^{2}$Mullard Space Science Laboratory, University College London, Dorking, RH5 6NT, UK\\
$^{3}$ Department of Physics, Columbia University, New York, NY 10027, USA\\
$^{4}$ Department of Physics and Astronomy, University of Pennsylvania, Philadelphia, PA 19104, USA\\
$^{5}$ School of Physics and Astronomy, Cardiff University, The Parade, Cardiff, Wales CF24 3AA, UK\\
$^{6}$ Department of Physics, Jadwin Hall, Princeton University, Princeton, NJ 08544, USA\\
$^{7}$ Department of Astrophysical Sciences, Peyton Hall, Princeton University, Princeton, NJ 08544, USA}

\date{Accepted XXX. Received YYY; in original form ZZZ}

\pubyear{2023}

\begin{document}
\label{firstpage}
\pagerange{\pageref{firstpage}--\pageref{lastpage}}
\maketitle

\begin{abstract}

We use the emulation framework \textsc{CosmoPower} to construct and publicly release neural network emulators of cosmological observables, including the Cosmic Microwave Background (CMB) temperature and polarization power spectra, matter power spectrum, distance-redshift relation, baryon acoustic oscillation (BAO) and redshift-space distortion (RSD) observables, and derived parameters. We train our emulators on Einstein-Boltzmann calculations obtained with high-precision numerical convergence settings, for a wide range of cosmological models including $\Lambda$CDM, $w$CDM, $\Lambda$CDM+$N_\mathrm{eff}$, and $\Lambda$CDM+$\Sigma m_\nu$. Our CMB emulators are accurate to better than 0.5\%  out to $\ell=10^4$ which is sufficient for Stage-IV data analysis, and our $P(k)$ emulators reach the same accuracy level out to $k=50 \,\, \mathrm{Mpc}^{-1}$, which is sufficient for Stage-III data analysis. We release the emulators via an online repository (\href{https://github.com/cosmopower-organization}{CosmoPower Organisation}), which will be continually updated with additional extended cosmological models. Our emulators accelerate cosmological data analysis by orders of magnitude, enabling  cosmological parameter extraction analyses, using current survey data, to be performed on a laptop. We validate our emulators by comparing them to \textsc{class} and \textsc{camb} and by reproducing cosmological parameter constraints derived from \emph{Planck} TT, TE, EE, and CMB lensing data, as well as from the Atacama Cosmology Telescope Data Release 4 CMB data, Dark Energy Survey Year-1 galaxy lensing and clustering data, and Baryon Oscillation Spectroscopic Survey Data Release 12 BAO and RSD data.
\end{abstract}




\section{Introduction}
Next-generation Cosmic Microwave Background (CMB) surveys, such as the Simons Observatory~\citep[SO;][]{Ade_2019}, CMB-S4~\citep{CMBS4_DSR,Abazajian_2022}, and CMB-HD~\citep{hd}, will provide high signal-to-noise measurements of the CMB temperature and polarization anisotropy down to small scales, characterized by their power spectra. The final seasons of the Atacama Cosmology Telescope~\citep[ACT;][]{Henderson:2015nzj} and of the South Pole Telescope~\citep[SPT;][]{2014SPIE.9153E..1PB} will also make progress in that direction in the forthcoming years. In addition, current and next-generation galaxy surveys including the Dark Energy Survey \citep[DES;][]{2016MNRAS.460.1270D}, the Kilo-Degree Survey \citep[KiDS;][]{2015MNRAS.454.3500K}, the Hyper Suprime-Cam Survey \citep[HSC;][]{2018PASJ...70S...4A}, the Vera C.~Rubin Observatory \citep{LSST:2008ijt}, \emph{Euclid} \citep[][]{EUCLID:2011zbd}, \emph{Roman} \citep{2015arXiv150303757S,2019arXiv190205569A} or \emph{SPHEREx} \citep{Dore:2014cca} probe the matter power spectrum via weak gravitational lensing and galaxy clustering, and spectroscopic galaxy surveys such as the Baryon Oscillation Spectroscopic Survey \citep[BOSS;][]{2013AJ....145...10D} and the Dark Energy Spectroscopic Instrument \citep[DESI;][]{DESI:2016fyo} also probe the baryon acoustic oscillation (BAO) feature and cosmological velocities via redshift-space distortions (RSD).

Einstein-Boltzmann codes such as \textsc{camb} \citep{camb2000ApJ...538..473L} and \textsc{class} \citep{classI,classII} are routinely used to accurately compute linear-theory cosmological power spectra, as well as the background cosmic evolution (\textit{e.g.}, the distance-redshift relation). These codes also implement various prescriptions for the calculation of non-linear corrections to the matter power spectrum and quantities derived therefrom, \textit{e.g.}, Halofit \citep{Smith_2003} and HMCode \citep{Mead15}. The computation of these cosmological quantities through a Boltzmann code represents a significant computational requirement in Markov Chain Monte Carlo (MCMC) methods, traditionally used to extract constraints on cosmological parameters. MCMC methods require $\sim 10^4 - 10^6$ calls to a Boltzmann code as they achieve convergence through a similar number of likelihood evaluations. The problem is exacerbated if one requires high numerical accuracy in the computation of the cosmological quantities obtained from Boltzmann codes, as needed for upcoming surveys~\citep[\textit{e.g.},][]{McCarthy:2021lfp}.

Several groups have developed emulators of cosmological quantities in order to accelerate the MCMC calculations by bypassing the call to a Boltzmann code with a faster algorithm \citep[\textit{e.g.}][]{Fendt07, Auld07, Albers:2019rzt, Mootoovaloo21, Arico21, Gunther:2022pto}. Recently, \cite{Spurio_Mancini_2022} developed \textsc{CosmoPower}, an emulation framework for cosmological quantities based on TensorFlow neural networks \citep{tensorflow2015-whitepaper}. The accuracy of the \textsc{CosmoPower} emulators presented in \citet{Spurio_Mancini_2022} was tested on CMB and large-scale structure analysis of current and future data. More recently, \cite{SPT-3G:2022hvq} constructed \textsc{CosmoPower} emulators for the analysis of the SPT data, covering angular scales up to $\ell=3000$.

Here we use \textsc{CosmoPower} to provide high-accuracy emulators of CMB temperature (emulator acronym: TT), polarization (TE and EE), and lensing potential power spectra (PP), up to $\ell_\mathrm{max}=10^4$, as well as linear and non-linear matter power spectra (PKL and PKNL) up to $k=50\,\mathrm{Mpc}^{-1}$ in the redshift range $z\in[0,5]$, the Hubble parameter (H), angular diameter distance (DA), $\sigma_8(z)$ (S8) for $z\in[0,20]$, and various derived parameters (DER) (see Sec.~\ref{sec:method} for a complete list).

With this suite of emulators, we are able to accurately reproduce marginalized posterior distributions from current CMB and large-scale structure (LSS) likelihood codes (based on CMB, matter power spectra, and BAO distances) within minutes on a laptop, whereas the original MCMC analyses took hours to days on large computing clusters. To accomplish this, we use the \textsc{cobaya} MCMC sampler and its implemented likelihoods \citep[][]{Torrado:2020dgo}. Our \textsc{CosmoPower} emulators are wrapped into \textsc{cobaya} via a simple wrapper that will be available online\footnote{The wrapper will be shared in a forthcoming paper which will cover in detail how to embed these emulators in future experiments' likelihood pipelines\label{fn:cppy}.}. We reproduce parameter constraints from the ACT Data Release 4~\citep{ACT:2020gnv,ACT:2020frw}, \emph{Planck} 2018 temperature, polarization, and lensing potential data \citep{Planck:2018lbu,plc_2020}, DES-Y1 \citep{DES:2017tss,Troxel_2018,Abbott_2018}, and BAO distances from BOSS and other surveys~\citep{Beutler2011MNRAS.416.3017B,Ross:2014qpa,Alam_2017}.

Our emulators are explicitly constructed with sufficient accuracy to remain valid for use in upcoming CMB analyses for the foreseeable future, including ACT, SPT, SO, CMB-S4, and CMB-HD, as well as ongoing galaxy surveys.

In addition to the emulators, we release the full pipeline to generate them, which can be used by the community to build emulators for other extensions to the standard cosmological model. The cosmological models covered here include
 $\Lambda$ Cold Dark Matter (CDM), $w$CDM, where $w$ is the dark energy equation of state; $\Lambda$CDM with varying $N_\mathrm{eff}$, where $N_{\rm eff}$ is the effective number of relativistic species at recombination; and $\Lambda$CDM with varying neutrino mass $\Sigma m_\nu$. In  $\Lambda$CDM, $\Lambda$CDM+$N_\mathrm{eff}$, and $w$CDM, we assume one massive and two massless neutrino states, as in the baseline \emph{Planck} analyses~\citep{plc_2020}. In $\Lambda$CDM+$\Sigma m_\nu$, we generate emulators for both the case of three species of neutrinos with degenerate mass, and the case of one massive and two massless neutrino states.

This paper is organized as follows. In Section \ref{sec:method} we outline our methodology, while in Section \ref{sec:boltacc} we study and calibrate accuracy settings of Boltzmann solvers by comparing \textsc{camb} and \textsc{class}. This sets the accuracy target for our emulators, which we validate in Section \ref{sec:emulacc}. In Section \ref{sec:fastan} we illustrate the validity and applicability of our emulators on a variety of likelihoods mentioned above. We conclude in Section \ref{sec:end}.

\section{Method}\label{sec:method}

The general methodology used to construct training and testing sets for our neural network emulators follows that of \cite{Spurio_Mancini_2022}. Here we provide a brief summary and refer the reader to that paper for further details. To compute all of the cosmological quantities on which our emulators are trained and tested we use \textsc{class} v2.9.4.\footnote{We initially attempted to use the latest \textsc{class} version, \textit{i.e.}, v3; however, it appears that the numerical accuracy of the code has changed compared to v2.9.4 and it was not clear how to recover sufficiently high accuracy for our purposes at high $\ell$ (see \href{https://github.com/lesgourg/class\_public/issues/494}{https://github.com/lesgourg/class\_public/issues/494}).  Thus, here we opted for v2.9.4. We note that the updates made in v3 compared to v2.9.4 are detailed on the repository webpage.}  We proceed in six steps, as follows.

\begin{enumerate}
    \item 
    We specify the cosmological parameters upon which the emulators are built, \textit{i.e.}, the inputs for our neural networks. We choose the following six $\Lambda$CDM parameters, as defined in \textsc{class}:
    \begin{itemize}
    \item $\omega_b$, the physical baryon density;
    \item $\omega_\mathrm{cdm}$, the physical dark matter density;
    \item $H_0$, the Hubble parameter today;
    \item $\tau$, the reionization optical depth;
    \item $n_\mathrm{s}$, the scalar spectral index;
    \item $\ln 10^{10}A_\mathrm{s}$; where $A_\mathrm{s}$ is the amplitude of primordial scalar perturbations.
    \end{itemize}
In extensions, we also add $w$, $N_\mathrm{eff}$, or $\Sigma m_\nu$, defined in the previous section.

Note that we opt for $H_0$ rather than for the angular size of the sound horizon at recombination, $100\theta_s$ (which we obtain as a derived parameter), due to the slightly different definitions of the angular acoustic scale implemented in \textsc{camb} and \textsc{class}\footnote{In \textsc{class}, $\theta_s$ is defined at the maximum of the visibility function, while in \textsc{camb} it is defined when the optical depth equals unity, and these times are slightly different. The $\theta_s$ of our emulator is the same as that in \textsc{class}.}. For MCMC analyses, if one needs to sample over $100\theta_s$ rather than $H_0$, it is straightforward to do so, for example using a shooting method with the derived parameter emulators\footnote{This adds an extra 20 ms to each MCMC step and would therefore increase slightly the convergence time of the chains. Note that the shooting method is also the standard way to sample over $100\theta_s$ in \textsc{class}.}.

    \item 
    We specify the range of the parameter values over which we train our emulators, reported in Table \ref{tab:priors}. These ranges should be conservative enough for all physically relevant analyses, assuming the use of data with similar or better constraining power than \emph{Planck} for the CMB, and DES or BOSS for galaxies.
    We emphasize that the emulators should not be used outside the ranges of Table \ref{tab:priors} or for different cosmological models, as there is no guarantee on their accuracy in such cases.
    For the $\Lambda$CDM+$N_\mathrm{eff}$  model, we let $N_\mathrm{eff}$ vary between $1.5$ and $5.5$. In practice, we vary the parameter $N_\mathrm{ur}$ in \textsc{class} between 0.49 and 4.49 and obtain $N_\mathrm{eff}$ as a derived parameter\footnote{$N_\mathrm{ur} = N_\mathrm{eff} - (T_{\rm ncdm}/(4/11)^{1/3})^4$, where $T_{\rm ncdm}$ is in units of $T_{\rm cmb}$ today. This is the convention used by \textsc{class}.}.
    In order to generate the data necessary for the matter power spectrum at different redshifts, we add $z_\mathrm{pk}$, the redshift at which we output the matter power spectra, as an extra varying input parameter to our grid.

    \item 
    We generate a Latin hypercube (LHC) of the parameter space. We set $N_\mathrm{S}=128,000$ as the number of samples in the LHC\footnote{This is roughly the same amount of samples as in a standard cosmological MCMC analysis.}, which is the same as the number of spectra we aim at computing. To generate the LHC, we use the python library \verb|pyDOE| \footnote{\href{https://pythonhosted.org/pyDOE/}{https://pythonhosted.org/pyDOE/}}.
    \item  We use a computing cluster to calculate the $N_\mathrm{S}$ training samples in parallel. Note that despite the total number of training samples being similar to that needed for an MCMC, the task of generating training samples is embarrassingly parallel, since the computation for a given set of input parameters does not depend on any other --- unlike an MCMC, which is a path-dependent calculation in parameter space. Given this, we find that  computing resources are more optimally used by running each computation on a single thread (\textit{i.e.}, setting \texttt{OMP\_NUM\_THREADS} = 1 before running \textsc{class}). The total amount of disk space taken by the training and testing samples is $\approx 150$GB. Note that for each sample, in addition to the cosmological observables that we want to emulate, we also save 14 derived parameters, such as $\sigma_8$ and $100\theta_\mathrm{s}$ (see the end of this section for details). For each class of emulators ($\Lambda$CDM + extensions) the computation of the data requires $\approx 48$ hours on a modern high-performance computing cluster. (The accuracy settings for the Boltzmann codes are given in Section \ref{sec:boltacc}.)

    \item   We process the data so that the training samples can be used in the \textsc{CosmoPower} training pipeline. For all quantities (TT, TE, EE, PP, PKL, PKNL, DA, H, S8, DER) we select 80\% of the samples for training and leave the remaining ones for testing. The TT, EE, PP, PKL, PKNL, H, and DER data consists of positive numbers with large dynamic ranges; to ease training, we take the logarithm of all these quantities before passing them to \textsc{CosmoPower}. The TE power spectra are oscillatory, zero-crossing functions, and thus we cannot directly take the logarithm. In this case, we use Principal Component Analysis (PCA), which reduces the dimensionality of the dataset and its dynamic range. We retain only 64 principal components for each spectrum. We verify that adding more does not lead to any significant improvement \citep{Spurio_Mancini_2022}.
    \item     We generate the emulators using the \textsc{CosmoPower} functions and verify their accuracy on the test set. Generating each emulator involves training a neural network model. This operation takes $\mathcal{O}(1\,\mathrm{hr})$ for each emulator and is best performed on a GPU, to fully take advantage of the acceleration provided by the \textsc{TensorFlow} library, which \textsc{CosmoPower} uses for its neural network implementations. We also stress that for a given cosmological model once the training is done and the emulator generated, this step does not need to be repeated. Note that we emulate TT, TE, EE, PP, PKL, PKNL, DA, H, S8, and DER separately.
    For all emulators we use dense neural networks with 4 hidden layers, each with 512 nodes. When building the emulators for TT, TE, EE, PP, DA, H, S8, and DER we do not include $z_{\rm pk}$ in the mapping from parameters to data, since $z_{\rm pk}$ has no effect on these quantities.  Similarly we also remove $\tau$ from the parameter sets for the PP, DA, H, S8, PKL, and PKNL emulators.

    The trained TT, TE, EE, PP, PKL, PKNL, DA, H, S8, and DER emulators are stored as \verb|pickle| files. The size of each CMB power spectra emulator file is $25.9$MB, except the TE emulator which is lighter because of the PCA and is $6.3$MB. The PKL and PKNL emulators are $4.2$MB; the H, DA, and S8 emulator files are $13.5$MB; and the DER emulator is $3.2$MB. (The size of each emulator file is proportional to the number of data points that are saved: 11000 multipoles for the CMB TT and EE spectra, the number of PCA weights for the TE spectra, 500 wavenumbers for PKL and PKNL, and 5000 redshifts for DA, H, and S8.)
\end{enumerate}

For illustration, examples of CMB power spectra  are shown in Figure~\ref{fig:spectras_vs_camb}; examples of matter power spectra calculations are shown in Figure~\ref{fig:mpks}; and examples of calculations of $H(z)$, $D_A(z)$, and $\sigma_8(z)$, which are used in the BAO and RSD calculations, are shown in Figure~\ref{fig:dahs8}.

\subsection*{Matter power spectrum}

To construct the emulators for the linear power spectrum and its non-linear corrections, we add an extra input to the neural networks, namely, the redshift at which the power spectra are computed, $z_{\rm pk}$. This represents an additional parameter, sampled between 0 and 5. For instance, the LHC used for our $\Lambda$CDM emulator has seven dimensions: the six $\Lambda$CDM parameters (cf. Table \ref{tab:priors}) augmented by $z_{\rm pk}$. For each LHC sample we save the matter power spectra at $z_{\rm pk}$ on a logarithmically spaced $k$-grid between $k_\mathrm{min}=10^{-4}\,\,\mathrm{Mpc}^{-1}$ and $k_\mathrm{max}=50\,\,\mathrm{Mpc}^{-1}$  with 500 points. (See Section \ref{sec:boltacc} for our perturbation and non-linear settings.)

\subsection*{BAO and redshift-space distortions}

Recent galaxy survey data allow us to constrain models based on various BAO distance measures, as well as using redshift-space distortions to constrain the parameter combination $f\sigma_8$, where $f$ is the growth rate of cosmic structures. To compute BAO distances and $f\sigma_8$, we need to save the angular diameter distance $D_A$, the Hubble parameter $H$, and $\sigma_8$ as a function of $z$, as well as the comoving sound horizon at baryon drag $r_d$, which is last in the list of derived parameters discussed above. We save $D_A(z)$, $H(z)$, and $\sigma_8(z)$ on a linearly spaced $z$-grid between $z_\mathrm{min}=0$ and $z_\mathrm{max}=20$ with 5000 points. The upper bound $z_\mathrm{max}=20$ is much higher than what is relevant to galaxy surveys; however we record the high-$z$ distances as they may be useful to other applications, such as studies of reionization or cosmic-dawn 21cm measurements.
With these quantities, we compute BAO distances and $f\sigma_8$ straightforwardly \citep[see, \textit{e.g.},][]{Alam_2017}. Note that $f\sigma_8=-(1+z)d\sigma_8(z)/dz$ where the derivative can be evaluated numerically. This is how we compute $f\sigma_8$, as is done in \textsc{class}.

\begin{table}
\begin{centering}
\begin{tabular}{|ccc|}
\hline
\hline
 Parameter & Min. Value & Max. Value \tabularnewline
\hline
\hline
$\ln10^{10}A_{{\rm s}}$ & 2.5 & 3.5 \tabularnewline
\hline
$\Omega_{{\rm cdm}}h^{2}$ & 0.08 & 0.20 \tabularnewline
\hline
$\Omega_{{\rm b}}h^{2}$ & 0.01933 & 0.02533 \tabularnewline
\hline
$H_{0}$ [km/s/Mpc] & 39.99 & 100.01 \tabularnewline
\hline
$n_{{\rm s}}$ & 0.8812(0.8) & 1.0492(1.2) \tabularnewline
\hline
$\tau$ & 0.02 & 0.12 \tabularnewline
\hline
\hline
$m_{\nu}$ [eV] & 0. & 2. \tabularnewline
\hline
\hline
$w$ & $-2.$ & $-0.33$ \tabularnewline
\hline
\hline
$N_{{\rm eff}}$ & 1.5 & 5.5 \tabularnewline
\hline
\hline

\end{tabular}

\par\end{centering}
\caption{  Parameter ranges used to generate the Latin hypercube of cosmological parameters used to compute the training data (see Section \ref{sec:method} for details). Our emulators should never be used outside these ranges. For the matter power spectrum emulators the LHC is supplemented by $z_{\rm pk}$, the redshift at which the power spectrum is evaluated, which is varied between 0 and 5. For $n_s$, the values in parentheses refer to the prior bounds that are used in the emulators for extensions, which are slightly broader than for the $\Lambda$CDM emulators (outside the parentheses).
}\label{tab:priors}

\end{table}

\subsection*{Derived parameters}\label{sec:derived_parameters}

We save 14 commonly used derived parameters, namely: \label{p:derlist}
\begin{itemize}
    \item the angular size of the sound horizon at decoupling $100\theta_\mathrm{s}$,
    \item the amplitude of matter clustering $\sigma_8$,
    \item the primordial Helium fraction $Y_{\rm P}$,
    \item the reionization redshift $z_\mathrm{reio}$,
    \item the number of effective relativistic degrees of freedom in the early Universe  $N_\mathrm{eff}$,
    \item the conformal time at which the visibility function reaches its maximum (\textit{i.e.}, the recombination time) $\tau_\mathrm{rec}$,
    \item its associated redshift $z_\mathrm{rec}$,
    \item the comoving sound horizon at recombination $r_\mathrm{s,rec}$,
    \item the conformal angular diameter distance to recombination $r_\mathrm{a,rec}$,
    \item the conformal time at which the photon optical depth crosses unity $\tau_\star$,
    \item the redshift  $z_\star$ at which the photon optical depth crosses unity,
    \item its associated comoving sound horizon $r_\mathrm{s,\star}$,
    \item and conformal angular diameter distance $r_\mathrm{a,\star}$
    \item and the comoving sound horizon at baryon drag $r_d$.
\end{itemize}
These parameters are saved into a list for each sample in the LHC, with the ordering given above.

\section{Boltzmann code Accuracy and settings}\label{sec:boltacc}

To ensure high precision in our calculations, we carefully study and calibrate the numerical precision parameters in \textsc{camb} and \textsc{class}.\footnote{See \citet{Lesgourgues_2011_CLASSIII} for calibration of the default numerical precision settings of \textsc{class}, which are determined to be sufficient for a \emph{Planck}-like experiment, but require to be adjusted for our purpose, as explained in this section.} For the maximum multipole, we choose $\ell_\mathrm{max}=11000$ in both codes. We chose $\ell_\mathrm{max}=11000$ as some forecasts and calculations, e.g., in the context of CMB-HD \citep{hd}, require  CMB power spectra at such high multipoles.
For non-linear corrections to the matter power spectrum, we use \textsc{HMcode} \citep{Mead_2021} with fiducial values $c_\mathrm{min}=3.13$ and $\eta_0=0.603$ in both codes. These values correspond to a dark matter-only scenario and are the same as in the baseline \emph{Planck} 2018 analyses~\citep{plc_2020}, justifying our choice.  For reionization modelling we kept the default \textsc{class} setting (i.e., the same as in \textsc{camb}).

We remark here that in future work it will be useful to train emulators in which these \textsc{HMcode} parameters are also varied, or for other non-linear regime prescriptions. Indeed, as shown by, \textit{e.g.}, \citet{McCarthy:2021lfp}, future CMB power spectrum data will be  sensitive to the impact of baryonic physics on the matter power spectrum (via CMB lensing); in addition, LSS statistics are sensitive to the nonlinear and baryonic prescriptions at high values of $k$.  With emulators that include these parameters, we will be able to marginalize over this uncertainty.

 The TT, TE, EE, and PP spectra computed with \textsc{camb}\footnote{We use \textsc{camb} v1.3.6.} are fully converged when using the following parameters \citep[see, \textit{e.g.},][]{McCarthy:2021lfp,Hill:2021yec}:

\noindent \verb|set_matter_power|:
\begin{itemize}
    \item \verb|kmax=10|
    \item \verb|k_per_logint=130|
    \item \verb|nonlinear=True|
\end{itemize}
\verb|set_for_lmax|:
\begin{itemize}
    \item \verb|lens_potential_accuracy=8|
    \item \verb|lens_margin=2050|
\end{itemize}
\verb|set_accuracy|:
\begin{itemize}
    \item \verb|AccuracyBoost=2.0|
    \item \verb|lSampleBoost=2.0|
    \item \verb|lAccuracyBoost=2.0|
    \item \verb|DoLateRadTruncation=False|
\end{itemize}
We use the \textsc{camb} spectra computed with these settings as a high-accuracy reference. This high-accuracy \textsc{camb} calculation takes $\mathcal{O}(1\,\mathrm{min})$ per sample on 16 threads.

It is possible to match \textsc{class} TT, TE, EE,  TE, PKL, and PKNL spectra to \textsc{camb} to better than 0.1\% precision (for $\ell<3000$ and $k<10\,\mathrm{Mpc}^{-1}$), by using the precision parameters in the \verb|cl_ref.pre| file in the \textsc{class} GitHub repository (this file ensures convergence at the 0.01\% level internally to \textsc{class} for TT and EE) \citep{Lesgourgues_2011_CLASSIII} and setting \verb|k_max_tau0_over_l_max: 15| to ensure high accuracy at $\ell>4000$ as discussed below). This is illustrated in Figures~\ref{fig:spectras_vs_camb_ratio_in_percents} and \ref{fig:matter_spectras_vs_camb_ratio_in_percents}, where this calculation is shown as the dotted-dashed yellow lines, labeled ``class prec''. We refer to it as the ultra-high-precision \textsc{class} prediction.
Nonetheless, this computation takes $\mathcal{O}(1\,\mathrm{h})$ which is prohibitively long to generate $\mathcal{O}(10^5)$ spectra for training the emulators.

By optimizing the tradeoff between precision and computation time, we find that the minimal settings required in order to match the high-precision \textsc{camb} calculation with \textsc{class} v2.9.4 are obtained by setting the following three parameters:\footnote{Note that these parameters are optimized here, whereas \citet{McCarthy:2021lfp} used extremely high-precision \textsc{class} parameters without optimizing the tradeoff between precision and computation time.}
\begin{itemize}
    \item  \verb|accurate_lensing: 1|
    \item  \verb|k_max_tau0_over_l_max: 15.|
    \item  \verb|perturb_sampling_stepsize: 0.05|
\end{itemize}
These are the precision parameters that we use in order to generate our emulator training data. The first parameter ensures that the lensed TT, TE, and EE spectra are converged for $\ell>3000$. Without it, they have non-physical oscillatory features. The second parameter ensures convergence at high-$\ell$ for all the spectra, including the unlensed ones. The third parameter is critical to get high-accuracy spectra over the whole multipole range. It is particularly important to get a converged PP spectra for $\ell<$ \verb|l_switch_limber| (default value in \textsc{class} v2.9.4: \verb|l_switch_limber=10|). Note that out of these three \textsc{class} parameters, only the second one has an impact on the linear and non-linear matter power spectra. This high-precision \textsc{class} calculation takes roughly the same amount of time as the \textsc{camb} one, \textit{i.e.}, $\mathcal{O}(1\,\mathrm{min})$.

\section{Emulators}\label{sec:emulacc}

Our trained TT, TE, EE, PP, PKL, PKNL, H, DA, S8, and DER emulators in $\Lambda$CDM, $w$CDM, $\Lambda$CDM+$N_\mathrm{eff}$, $\Lambda$CDM+$\Sigma m_\nu$ are made publicly available on the \textsc{CosmoPower} GitHub repository.\footnote{\href{https://github.com/cosmopower-organization}{https://github.com/cosmopower-organization}}  To predict the power spectra, they need to be imported in Python, via the \textsc{CosmoPower} package.  They allow for rapid computation of power spectra and derived parameters, and hence of parameter posterior probability distributions in MCMC analyses (see Section~\ref{sec:fastan}). Loading the emulators takes $\approx 0.1$ sec. Computing one set of TT, TE, EE, PKL, PKNL power spectra and derived parameters takes $\approx 60$ ms, compared to $\mathcal{O}(1\,\,\mathrm{min})$ with \verb|class| or \verb|camb| with high-precision settings, \textit{i.e.}, we achieve a factor of 1000 speed-up. The GitHub repository also contains a set of notebooks showing how to use the emulators.

 \begin{figure*}
    \includegraphics[width=2.\columnwidth]{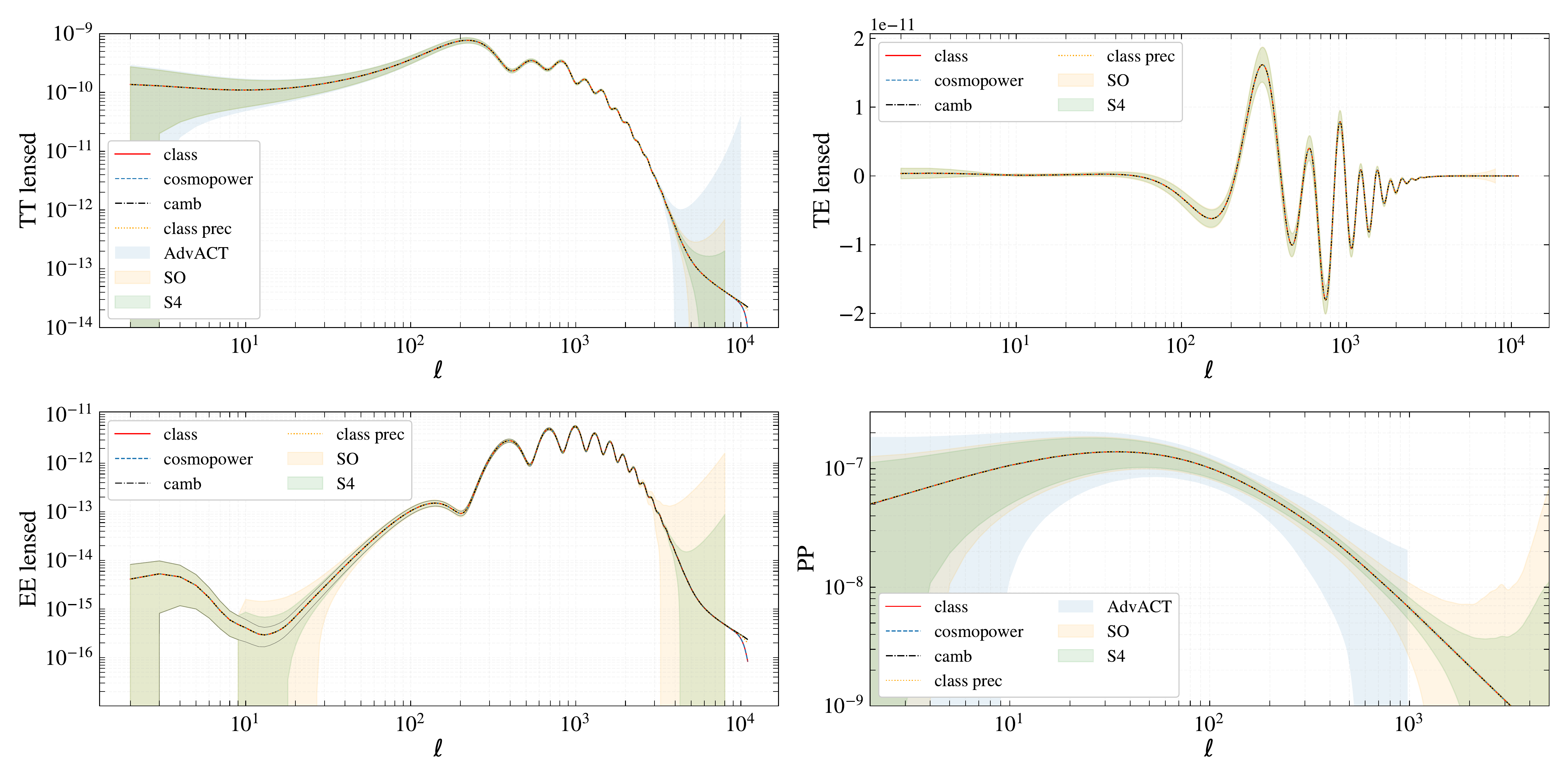}
    \vspace{0.2cm}
    \caption{Comparison of lensed CMB power spectra computed with \textsc{class} (for default precision as the solid red line and ultra-high precision settings as the dotted yellow line labeled "class-prec"), \textsc{CosmoPower}, and \textsc{camb}. We show the dimensionless $\ell(\ell+1)C_\ell/2\pi$ for TT, TE and EE, and $[\ell(\ell+1)]^2C_\ell^{\phi\phi}/2\pi$ for PP (lensing convergence power spectrum) in a $\Lambda$CDM model with one set of cosmological parameters from the \emph{Planck} 2018 results (see text for details). The shaded areas indicate the forecast 1$\sigma$ uncertainty for Advanced ACT, SO, and CMB-S4 (see Section \ref{sec:emulacc} for details). This figure illustrates the emulated quantities, the relative difference between the spectra are shown in Figure \ref{fig:spectras_vs_camb_ratio_in_percents} (in percent) and \ref{fig:spectras_vs_camb_ratio_in_sigmas4} (in units of CMB-S4 statistical error bars). On the EE plot, the thin black solid lines indicate the cosmic variance.}
    \label{fig:spectras_vs_camb}
\end{figure*}

  \begin{figure*}
    \includegraphics[width=2.\columnwidth]{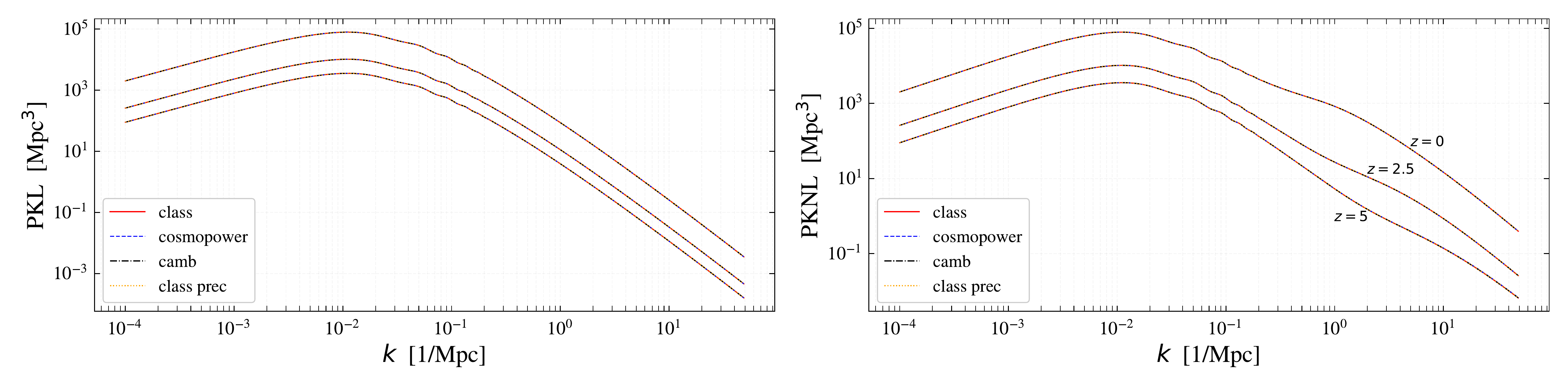}
    \vspace{0.2cm}
    \caption{ Same settings as Figure \ref{fig:spectras_vs_camb} but for matter power spectra. We show the linear matter power spectra on the left and the non-linear spectra on the right (computed according to \protect\texttt{HMcode} with $c_\mathrm{min} = 3.13$ and $\eta_0 = 0.603$, within \protect\textsc{class}). In both panels, we show the power at three redshifts, $z=0,2.5,5$ from top to bottom, spanning the range of the emulators. This figure illustrates the emulated quantities, the relative difference between these spectra are shown in percentages in Figure~\ref{fig:matter_spectras_vs_camb_ratio_in_percents}.}
    \label{fig:mpks}
\end{figure*}

  \begin{figure*}
    \includegraphics[width=2.\columnwidth]{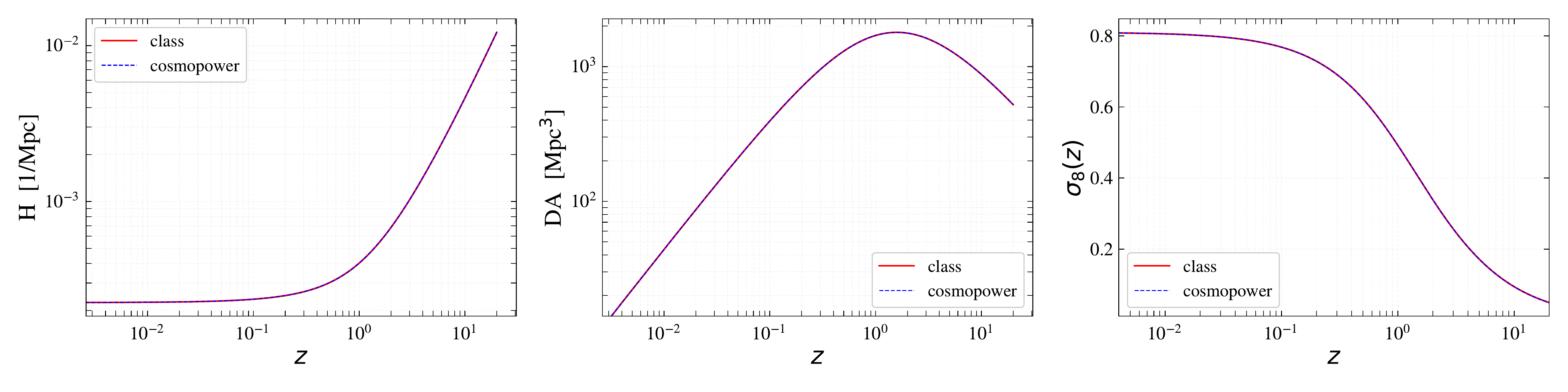}
    \vspace{0.2cm}
    \caption{Redshift evolution of the Hubble parameter (left), angular diameter distance (middle), and $\sigma_8$ (right) as computed with \textsc{class} and \textsc{CosmoPower}, between $z=0$ and 20 (\textit{i.e.}, spanning the redshift range of our emulators). We use the same settings as Figure \ref{fig:spectras_vs_camb} and note that in this parameters configuration the relative difference between \textsc{class} and \textsc{CosmoPower} (not shown here) is less than 0.2\%. This figure illustrates the emulated quantities, see Figure \ref{sec:emulacc} for the relative difference over the full testing set.}
    \label{fig:dahs8}
\end{figure*}

\begin{figure*}
    \includegraphics[width=2.\columnwidth]{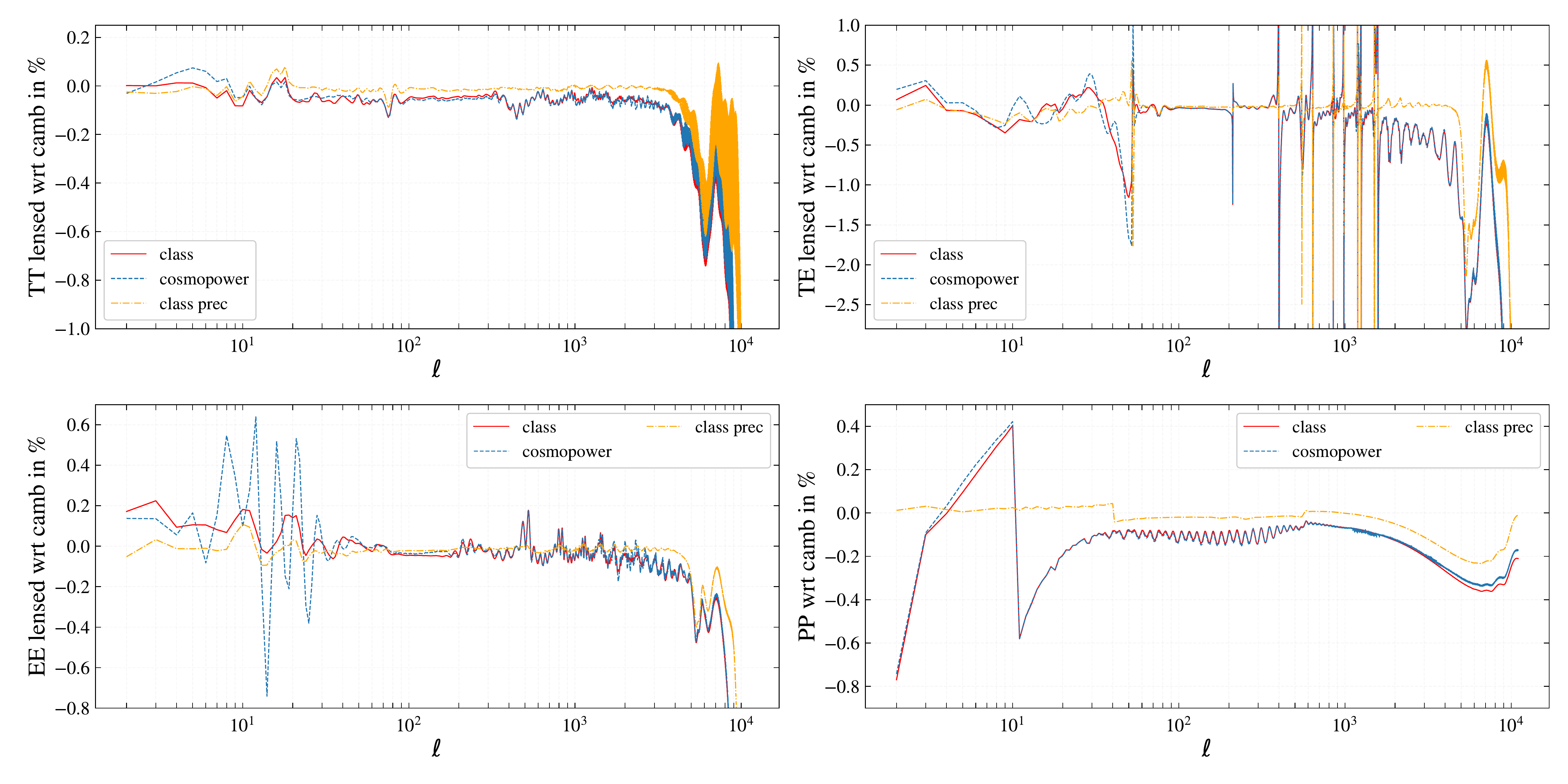}
    \vspace{0.2cm}
    \caption{ Relative difference of CMB power spectra (top left: TT, top right: TE, bottom left: EE, bottom right: lensing potential, PP) between two different settings of \textsc{class} (high precision as the solid red line and ultra-high precision settings as the dotted yellow line labeled "class-prec"), and \textsc{CosmoPower} with respect to \textsc{camb} in percent. We use the same parameters settings as in Figure \ref{fig:spectras_vs_camb}.  Note that the non-negligible fractional deviations seen in a few places for TE occur due to the zero-crossing of the spectrum at those multipoles, and do not reflect inaccuracy in the emulator.
    }
    \label{fig:spectras_vs_camb_ratio_in_percents}
\end{figure*}

\begin{figure*}
    \includegraphics[width=2.\columnwidth]{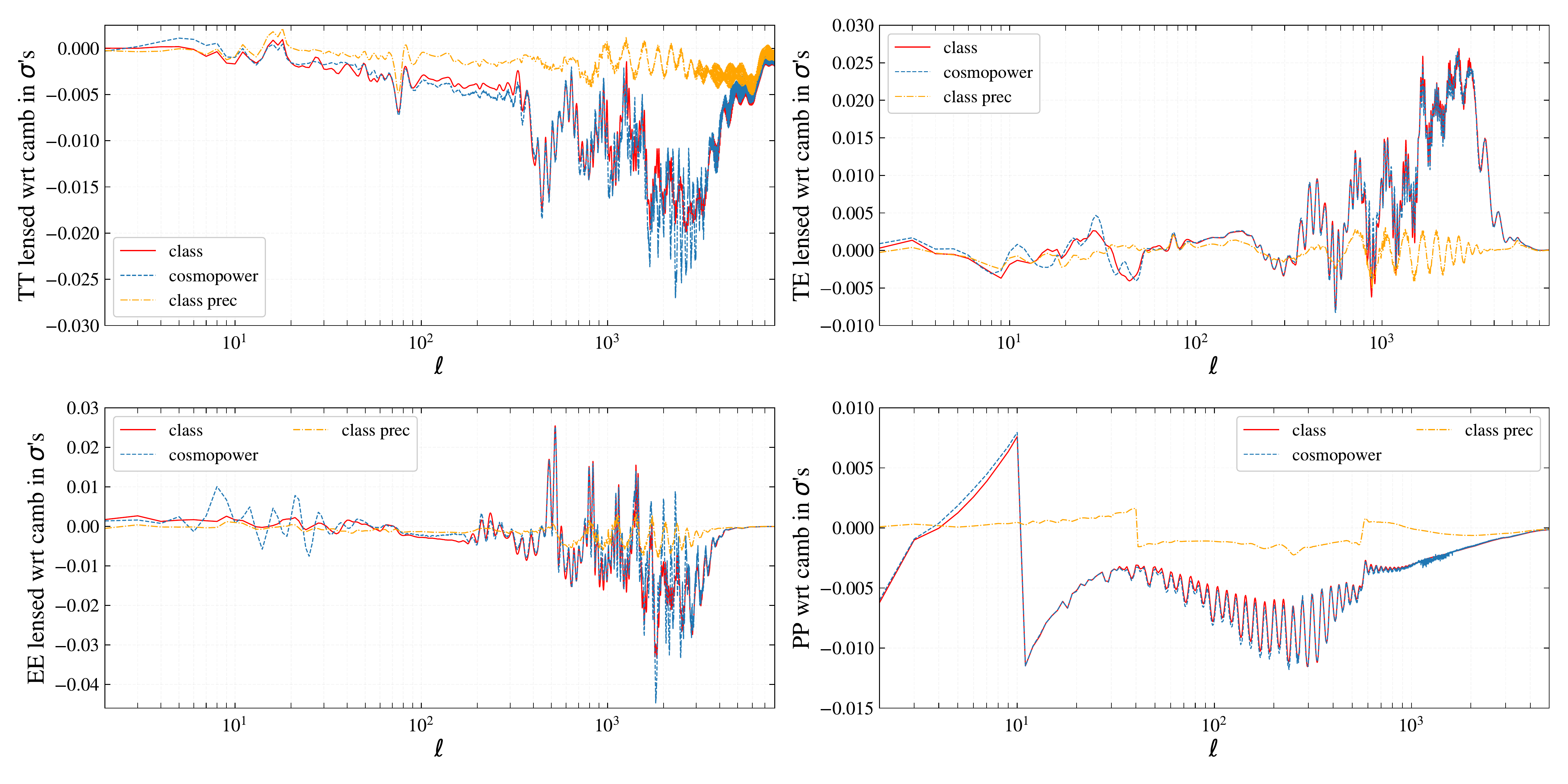}
    \vspace{0.2cm}
    \caption{Relative difference between two different settings of \textsc{class} (high precision as the solid red line and ultra-high precision settings as the dotted-dashed yellow line labeled "class-prec") and \textsc{CosmoPower} with respect to  \textsc{camb} in units of the forecast CMB-S4 statistical error bars.  Same settings as Figure \ref{fig:spectras_vs_camb}. }
    \label{fig:spectras_vs_camb_ratio_in_sigmas4}
\end{figure*}

\begin{figure*}
    \includegraphics[width=2.\columnwidth]{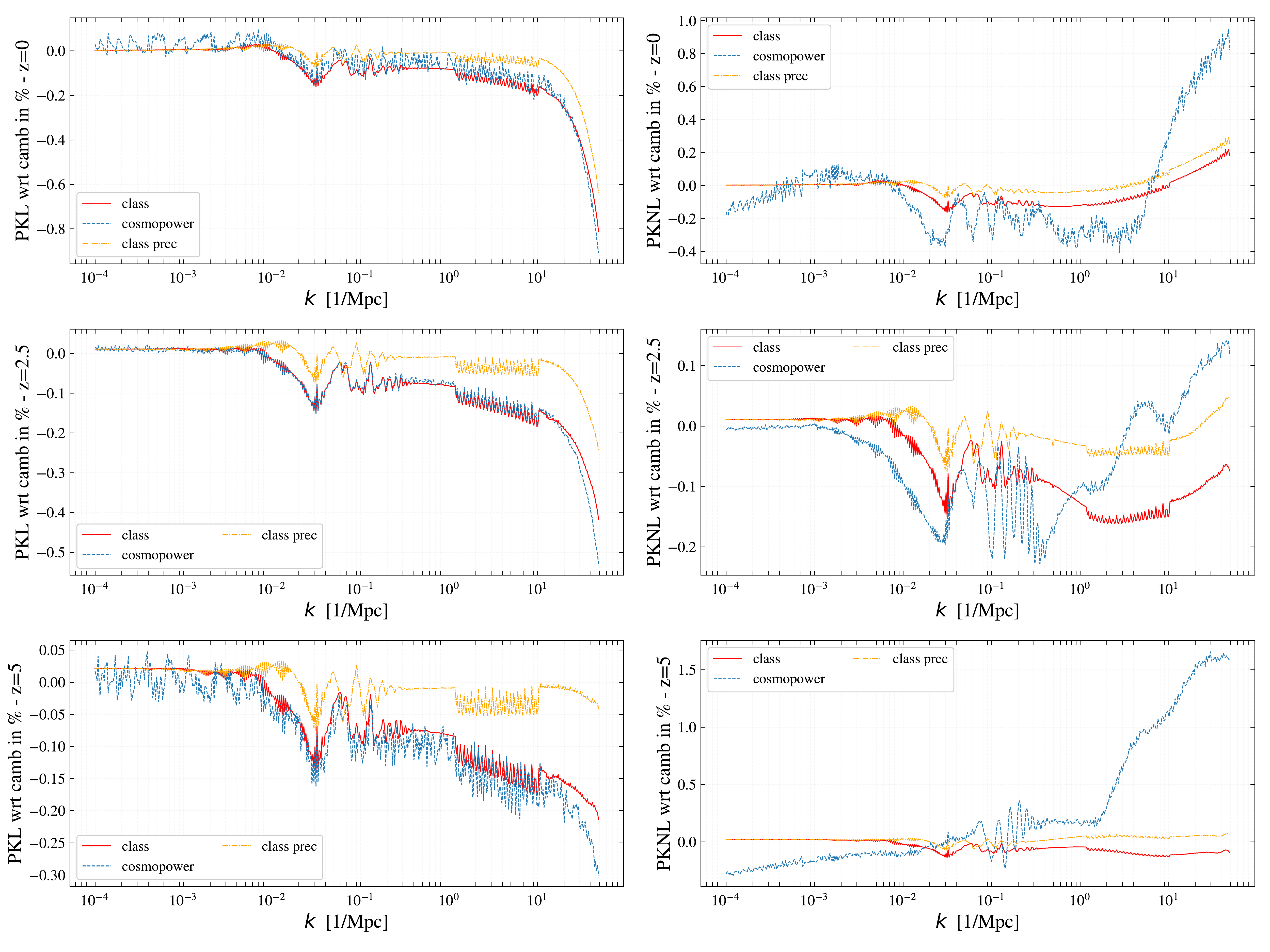}
    \vspace{0.2cm}
    \caption{ Relative difference of the linear (left) and non-linear (right) matter power spectra  between two different settings of \textsc{class} (high precision as the solid red line and ultra-high precision settings as the dotted-dashed yellow line labeled "class-prec")  and \textsc{CosmoPower} with respect to \textsc{camb} in percent.}
    \label{fig:matter_spectras_vs_camb_ratio_in_percents}
\end{figure*}

In Figure \ref{fig:spectras_vs_camb}-\ref{fig:matter_spectras_vs_camb_ratio_in_percents} we show the emulator's predictions for one set of cosmological parameters, namely, the central values of the right column of Table I in \protect\cite{plc_2020}, with one massive neutrino with $m_\nu = 0.06\,\mathrm{eV}$ and in $\Lambda$CDM.

In Figure \ref{fig:spectras_vs_camb}, we compare the power spectra from \textsc{class}, \textsc{camb}, and \textsc{CosmoPower} explicitly,   The \textsc{class} (\textit{i.e.}, the high-precision calculation discussed in Sec.~\ref{sec:boltacc}), ``class prec'' (\textit{i.e.}, the ultra-high-precision calculation discussed in Sec.~\ref{sec:boltacc}), high-precision \textsc{camb}, and \textsc{CosmoPower} power spectra are indistinguishable for $\ell<10000$.   For $10000<\ell<11000$, some differences can be seen in the TT and EE power spectra: the \textsc{class} and \textsc{CosmoPower} predictions tend to fall off compared to \textsc{camb}. Since this is less significant for the ``class prec'' prediction, we attribute this to numerical precision settings, and do not investigate further as this is probably irrelevant for any CMB experiment in the foreseeable future. The shaded areas indicate the sensitivity of Advanced ACT \citep{Henderson:2015nzj}, SO \citep{Ade_2019}, and CMB-S4 \citep{Abazajian_2022}  forecast sensitivity. We refer to Section 4 of \cite{Bolliet:2022pze} for details on the TT and PP sensitivity curves and the links where they can be downloaded from.  The TE sensitivity can be computed from TT and EE\footnote{The CMB-S4 EE noise is from  \texttt{S4\_190604d\_2LAT \_pol\_default\_noisecurves\_deproj0\_SENS0 \_mask\_16000\_ell\_EE\_BB.txt} and the SO EE noise is from \texttt{v3.1.0/SO\_LAT\_Nell\_P\_baseline\_fsky0p4\_ILC\_CMB\_E.txt}} \citep[see, e.g.,][ for the formula we use]{Spurio_Mancini_2022}. Note that we use $f_\mathrm{sky}=0.3$ for Advanced ACT and $f_\mathrm{sky}=0.4$ for SO and CMB-S4.  Also, note that all noise curves also assume the use of \emph{Planck} data in combination with that from each ground-based experiment (see Sec.~2 of~\citet{Ade_2019} for details); the use of \emph{Planck} is important in extending the forecast measurements to low multipoles, where otherwise atmospheric noise in the ground-based data would be very large.

Figure \ref{fig:mpks} shows the same as Figure~\ref{fig:spectras_vs_camb} but for the linear and non-linear matter power spectra, computed at three redshifts ($z=0,2.5,5$). In these cases, the \textsc{class}, \textsc{camb}, and \textsc{CosmoPower} predictions are indistinguishable across the full $k$ range. The same is true for the H, DA, and S8 predictions, as can be seen in Figure \ref{fig:dahs8}.

In Figure \ref{fig:spectras_vs_camb_ratio_in_percents}, we show the relative difference between CMB angular power spectra (TT,TE,EE,PP) and \textsc{camb}, in percent. The spikes in TE are where the spectra cross zero and are not problematic. As discussed before, the agreement between the ultra-high precision \textsc{class} prediction and \textsc{camb} is at the 0.1\% level until $\ell\sim 3000$, and degrades at higher $\ell$. Nonetheless, the agreement between \textsc{CosmoPower} and both \textsc{class} predictions remains at the 0.1\% level across the whole range of multipoles, including $\ell>3000$, except in two cases. First, in the low-$\ell$ part ($\ell<10$) of the lensing convergence power spectrum we see very small differences ($\lesssim 0.5$\%) between the ultra-high precision \textsc{class} prediction on one hand and the high-precision and \textsc{CosmoPower} prediction on the other hand (the agreement between the \textsc{CosmoPower} and high-precision \textsc{class} calculation remains better than 0.005\%). This difference can be reduced further by decreasing the parameter \verb|perturb_sampling_stepsize| (see Sec \ref{sec:boltacc}), at the cost of a longer runtime. Second, in the low-$\ell$ part ($\ell<50$) of the EE power spectrum we see differences between the \textsc{CosmoPower} and high-precision \textsc{class} prediction at the 0.6\% level. If needed, this could be solved by generating more training data.

To assess whether this level of difference is acceptable for CMB Stage-IV analyses, we compare the relative difference between the CMB spectra in terms of CMB-S4 sensitivity in Figure \ref{fig:spectras_vs_camb_ratio_in_sigmas4}. We see that the \textsc{cosmopower} power spectra agree with \textsc{camb} to better than $0.03\sigma$ across all multipoles and with the high-precision \textsc{class} prediction to better than $0.03\sigma$. Therefore, such  level of agreement is sufficient.

In Figure~\ref{fig:matter_spectras_vs_camb_ratio_in_percents} we show the relative difference between the linear and non-linear matter power spectra with respect to \textsc{camb} in percent, at three redshifts, $z=0,2.5$, and 5. In the left panels, we see that the \textsc{CosmoPower} linear matter power spectra agree with the high-precision \textsc{class} predictions at the 0.05\% level across the whole $k$ range. The difference between these and the ultra-high precision \textsc{class} prediction is at the 0.3\% level. Finally, the difference between the ultra-high precision \textsc{class} prediction and \textsc{camb} is at the 0.3\% level up to $k\approx20\,\mathrm{Mpc}^{-1}$ and then becomes $\approx 0.8\%$ at higher $k$. In the right panels, showing the non-linear matter power spectra, we see that differences between \textsc{cosmopower} on one hand and \textsc{class} and \textsc{camb} on the other hand are more important for the non-linear matter power spectra, becoming larger in the non-linear regime, but remain at the $1$-$1.5$\% level, including at $z=5$.

Since there is no direct measurement of the matter power spectrum we cannot compare it with an instrumental sensitivity level. However, we can check the accuracy on observables based on the matter power spectrum. The lensing convergence power spectrum (PP, bottom right panel of Figure \ref{fig:spectras_vs_camb}) is one such example. Other examples include galaxy clustering and galaxy weak lensing observables and to a lesser extent, the lensing of CMB angular power spectra. We perform Stage III posterior inference analyses based on these observables and recover nearly identical constraints to those obtained running the full Boltzmann code calculations (see Section \ref{sec:fastan}), giving us confidence that our emulators are sufficiently precise. We defer to future work a more detailed quantitative analysis, including the release of high-accuracy emulators for the matter power spectrum tested against (forecast) Stage-IV configurations.

\begin{figure*}
    \includegraphics[width=2.\columnwidth]{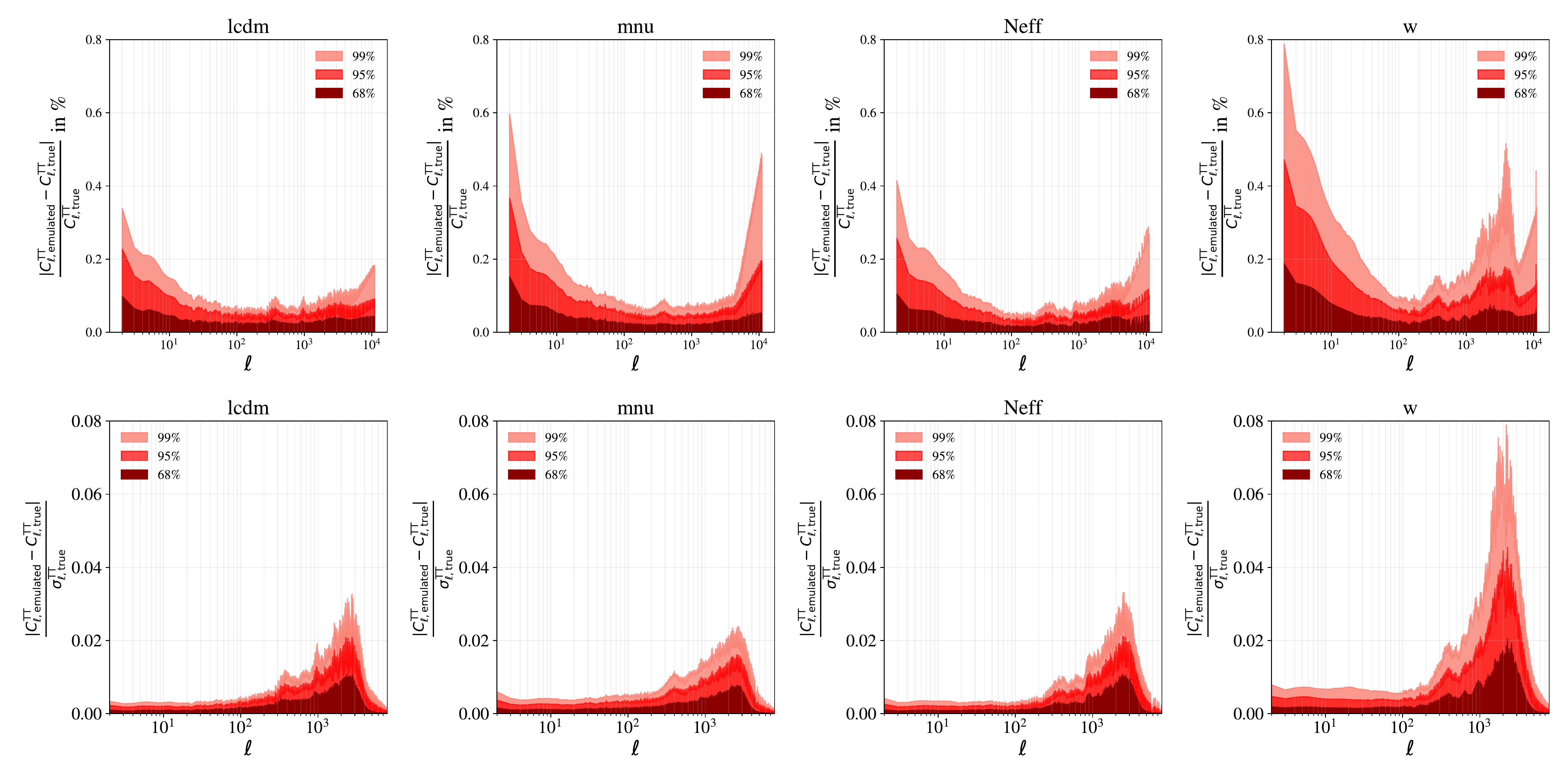}
    \vspace{0.2cm}
    \caption{ Relative difference between \textsc{CosmoPower} and the ``true'' high-precision \textsc{class} prediction for CMB TT  power spectra in percent (top) and in units of CMB-S4 sensitivity (bottom). See end of Section \ref{sec:emulacc} for details.}
    \label{fig:TT_spectras_pred_vs_truth}
\end{figure*}

\begin{figure*}
    \includegraphics[width=2.\columnwidth]{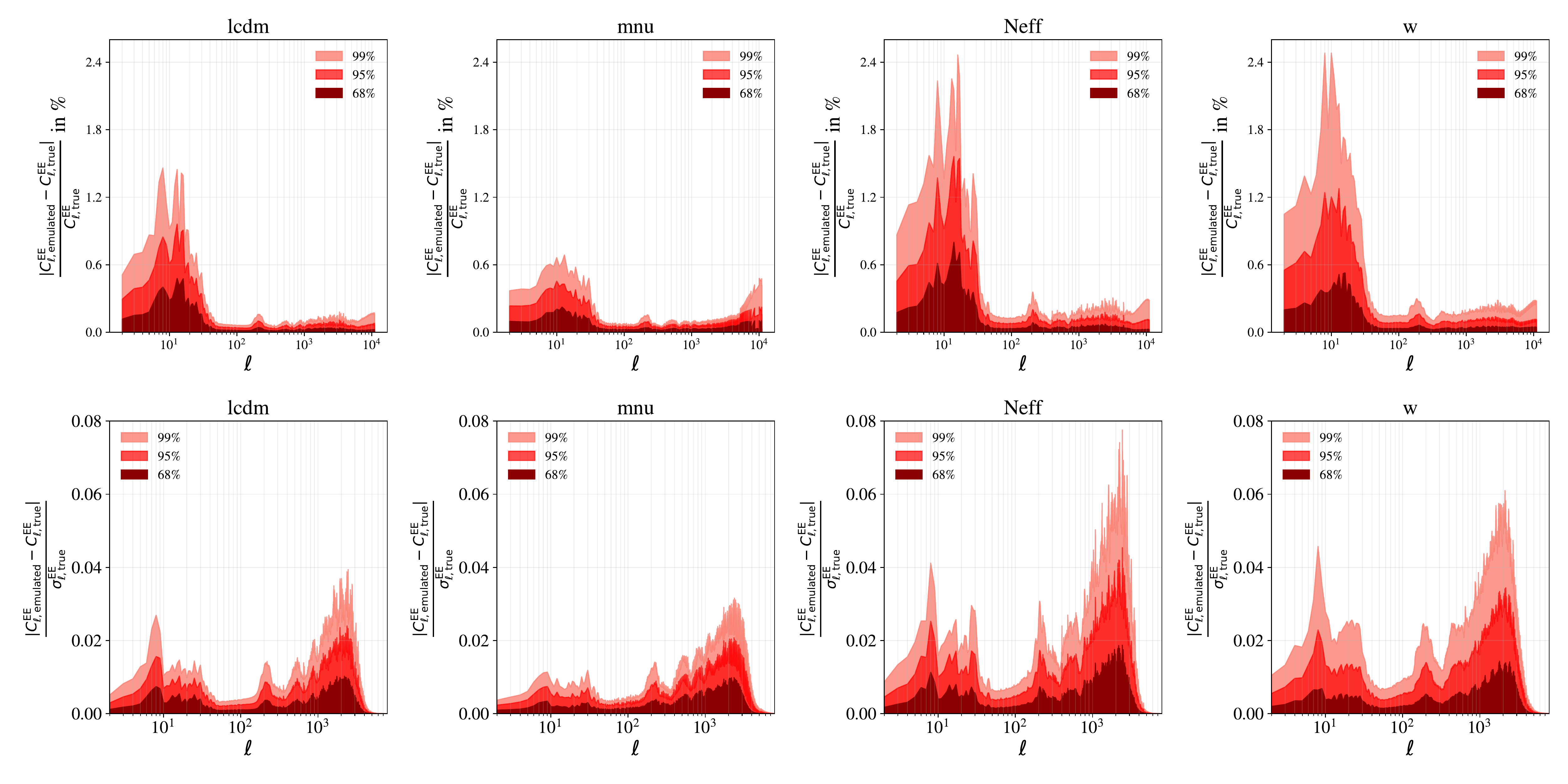}
    \vspace{0.2cm}
    \caption{ Relative difference between \textsc{CosmoPower} and the ``true'' high-precision \textsc{class} prediction for CMB EE power spectra in percent (top) and in units of CMB-S4 sensitivity (bottom). See end of Section \ref{sec:emulacc} for details.}
    \label{fig:EE_spectras_pred_vs_truth}
\end{figure*}

\begin{figure*}
    \includegraphics[width=2.\columnwidth]{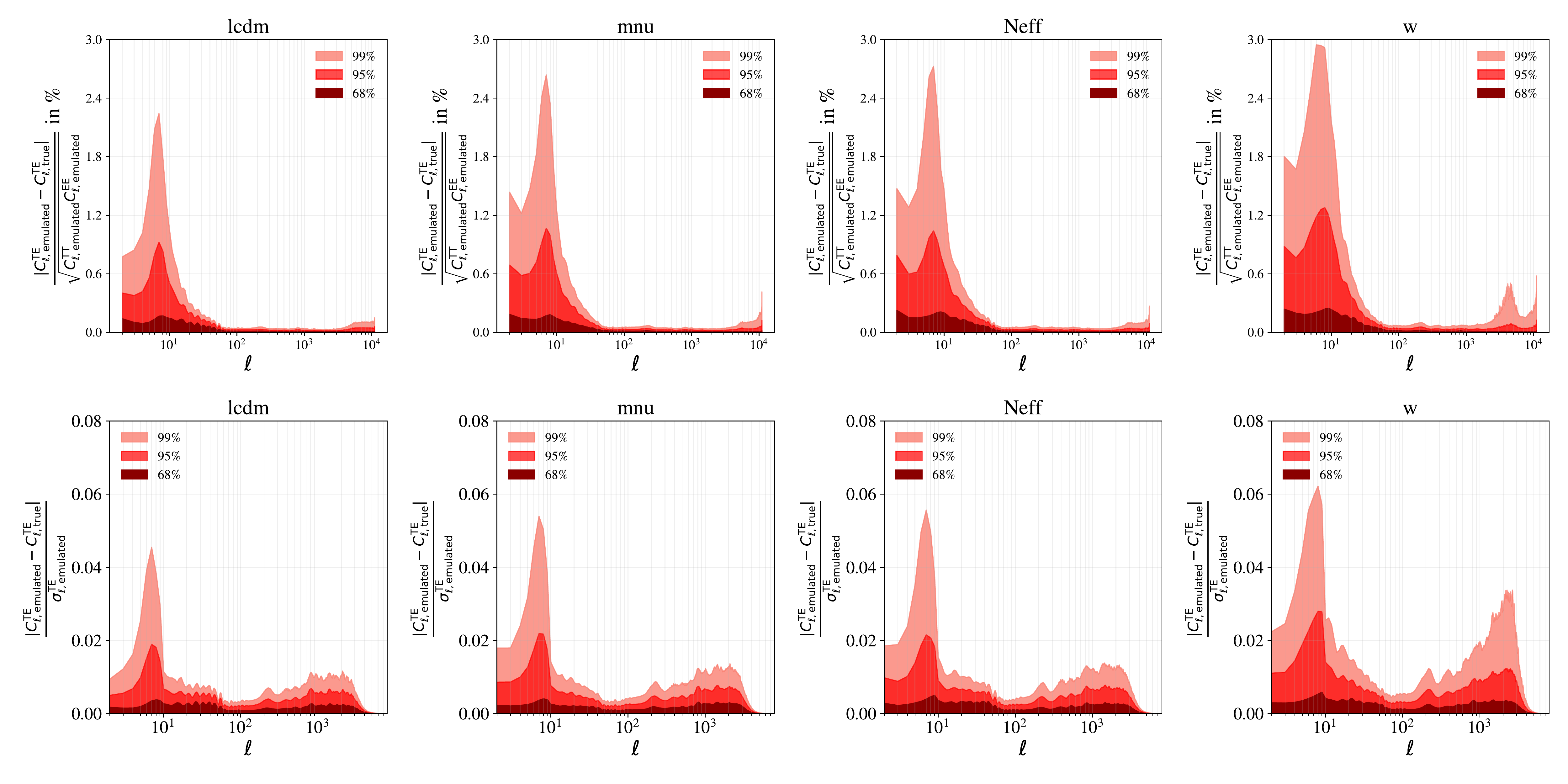}
    \vspace{0.2cm}
    \caption{ Relative difference between \textsc{CosmoPower} and the ``true'' high-precision \textsc{class} prediction for CMB TE power spectra in percent (top) and in units of CMB-S4 sensitivity (bottom).  Note that the spikes are simply due to zero crossings of the TE power spectrum. See end of Section \ref{sec:emulacc} for details.}
    \label{fig:TE_spectras_pred_vs_truth}
\end{figure*}

\begin{figure*}
    \includegraphics[width=2.\columnwidth]{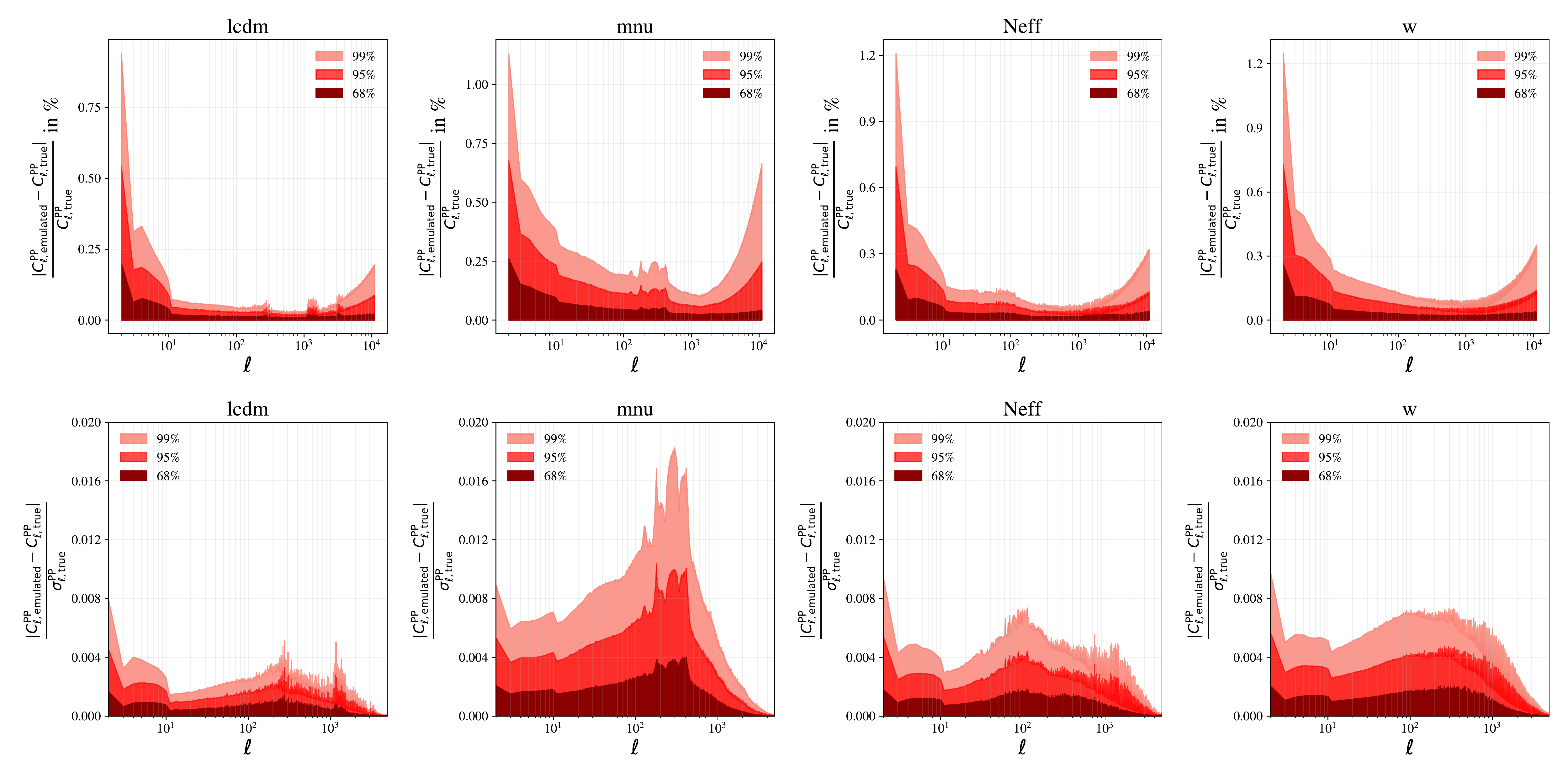}
    \vspace{0.2cm}
    \caption{ Relative difference between \textsc{CosmoPower} and the ``true'' high-precision \textsc{class} prediction for CMB lensing convergence  power spectra in percent (top) and in units of CMB-S4 sensitivity (bottom). See end of Section \ref{sec:emulacc} for details.}
    \label{fig:PP_spectras_pred_vs_truth}
\end{figure*}

\begin{figure*}
    \includegraphics[width=2.\columnwidth]{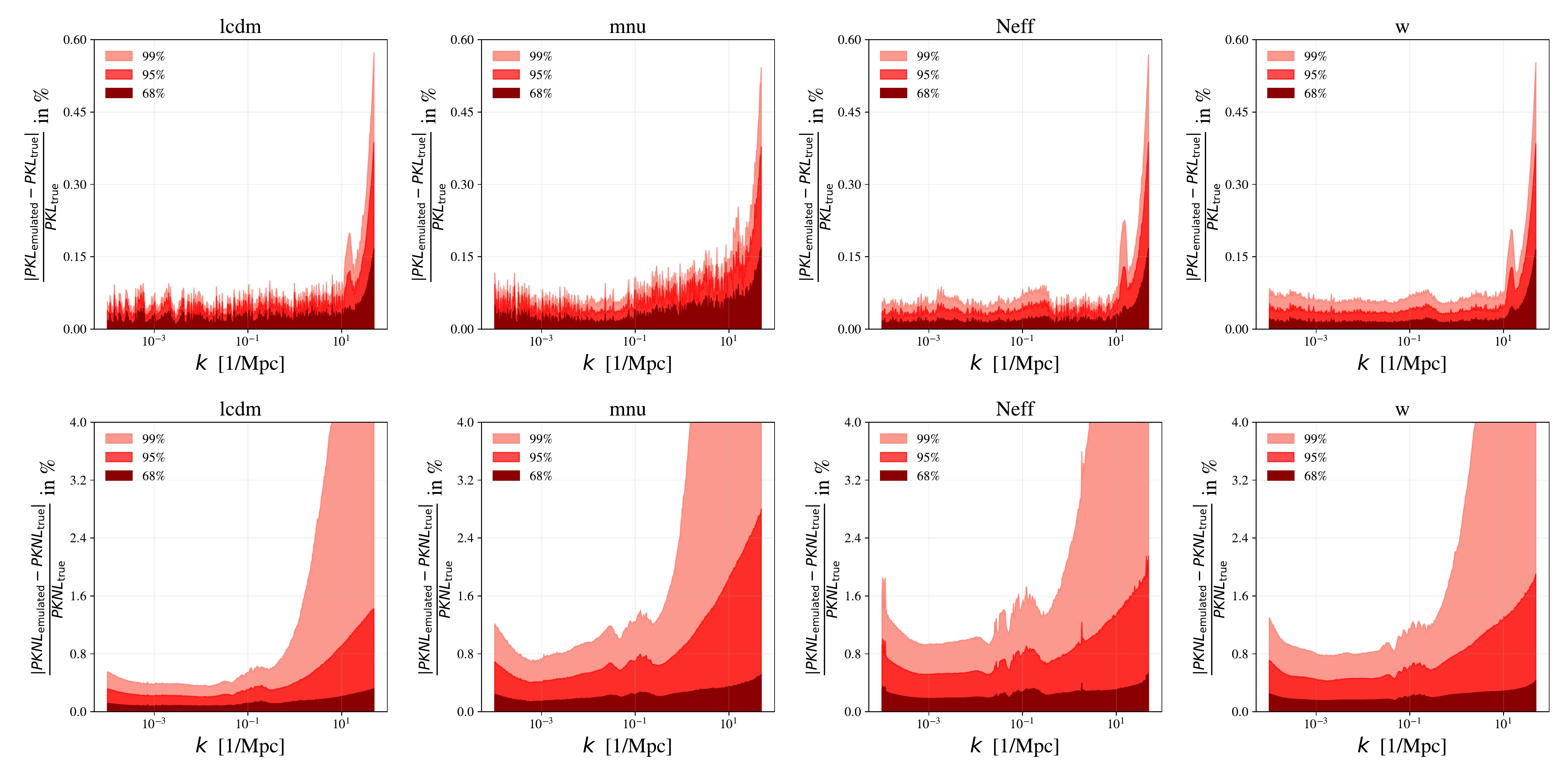}
    \vspace{0.2cm}
    \caption{ Relative difference between \textsc{CosmoPower} and the ``true'' high-precision \textsc{class} prediction for the linear (top) and non-linear (bottom) matter power spectra in percent. The \textsc{class} predictions used here are the test dataset, made of $\approx 25000$ spectra. See end of Section \ref{sec:emulacc} for details.
    }
\label{fig:PKLNL_pred_vs_truth_ratio_in_percents}
\end{figure*}

\begin{figure*}
    \includegraphics[width=2.\columnwidth]{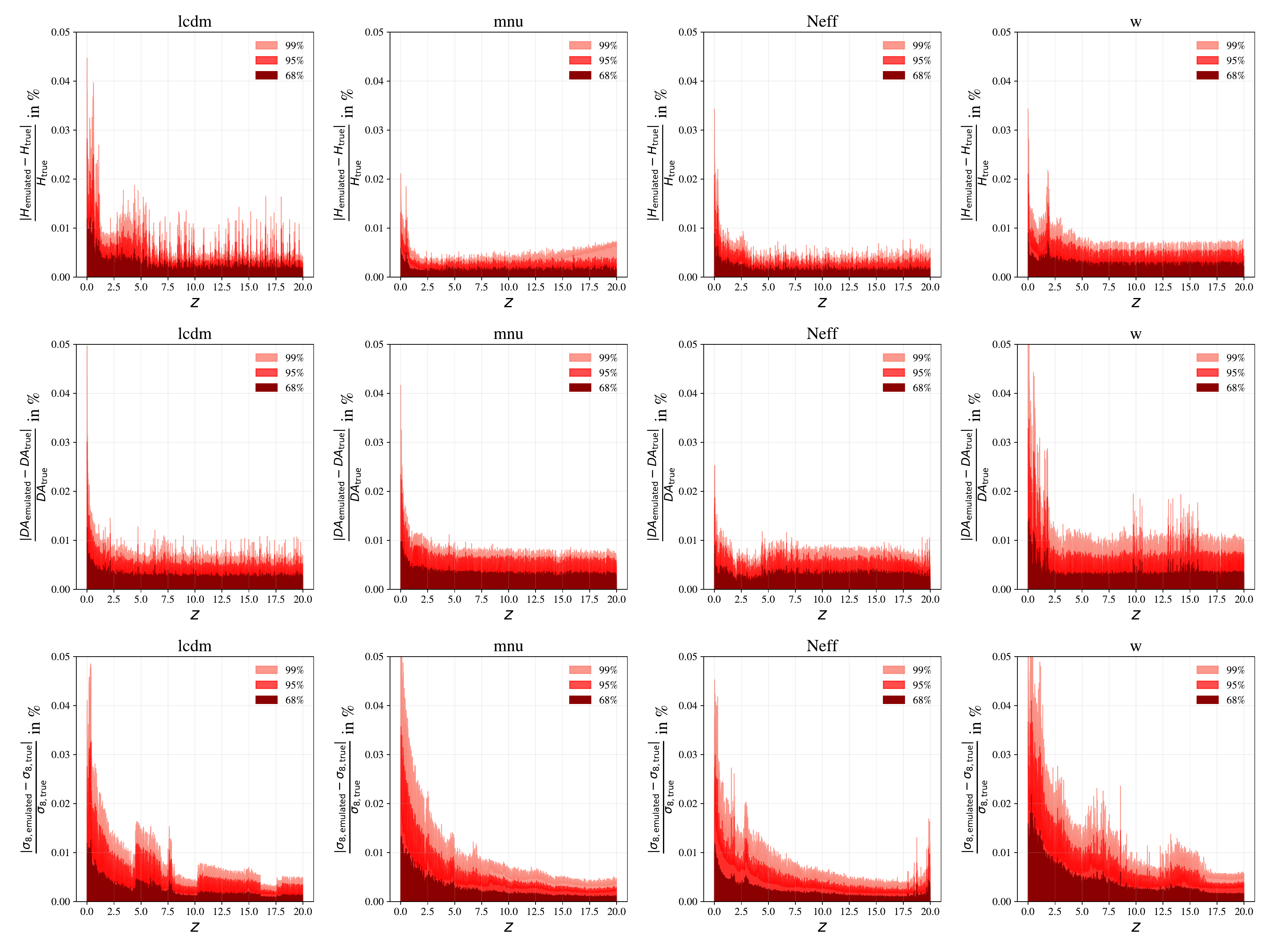}
    \vspace{0.2cm}
    \caption{
    Relative difference between \textsc{CosmoPower} and the ``true'' high-precision \textsc{class} prediction for the redshift evolution of the Hubble parameter (top row), angular diameter distance (middle row), and $\sigma_8$ (bottom row) in percent. See end of Section \ref{sec:emulacc} for details.}
    \label{fig:HDAs8_pred_vs_truth_ratio_in_percents}
\end{figure*}

\begin{figure*}
    \includegraphics[width=2.\columnwidth]{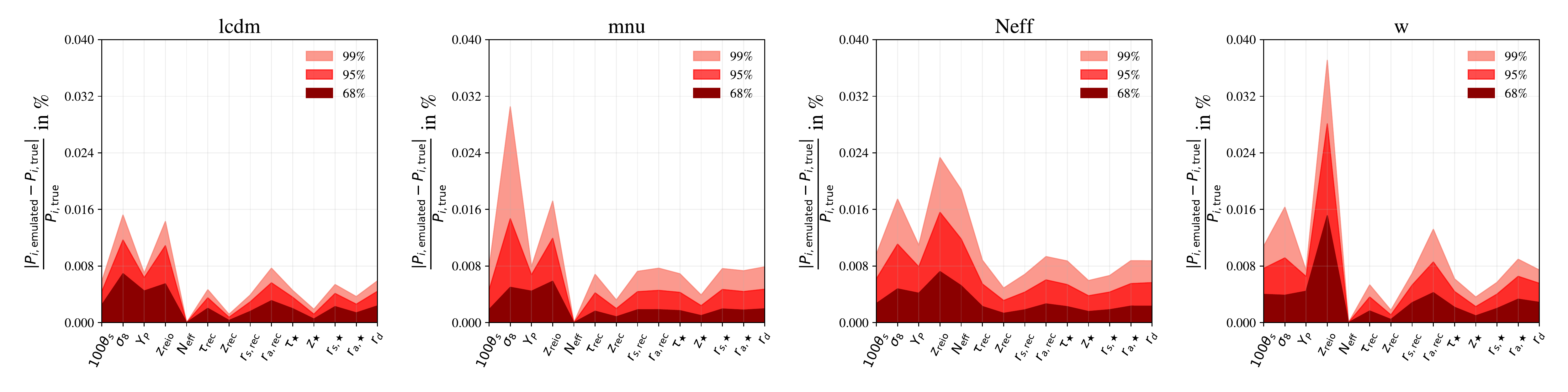}
    \vspace{0.2cm}
    \caption{Relative difference between \textsc{CosmoPower} emulators and the ``true'' high-precision \textsc{class} prediction for the derived parameters in percent. Each unit on the x-axis represents a derived parameter as indicated (see page \pageref{p:derlist} for the definitions of the fourteen parameters).  Note that for $N_{\rm eff}$ the error is null since this parameter is fixed, except in the $\Lambda$CDM+$N_\mathrm{eff}$ model (third panel) where it is varied. In $\Lambda$CDM+$N_\mathrm{eff}$, $N_\mathrm{eff}$ is not an input parameter but a derived parameter (see Section \ref{sec:method}) and hence there is a small error. }
    \label{fig:der_pred_vs_truth_ratio_in_percents}
\end{figure*}

The results shown in Figures~\ref{fig:spectras_vs_camb}-\ref{fig:matter_spectras_vs_camb_ratio_in_percents}, are for one set of cosmological parameters. However, it is important to quantify the precision of our emulators over the whole prior range of parameters (see Table \ref{tab:priors}), in all the cosmological models. We do so in Figures~\ref{fig:TT_spectras_pred_vs_truth}-\ref{fig:der_pred_vs_truth_ratio_in_percents}\footnote{The "mnu" results on this figure are for one massive and two massless neutrino states.}. These figures show the relative difference between the emulators, \textit{i.e.}, the \textsc{CosmoPower} prediction, and the exact high-precision \textsc{class} prediction on the testing data sets\footnote{The testing data is made of the exact computations which are not used for training the emulator neural networks. We use 20\% of our dataset for testing.}, containing $\approx 25000$ predictions. The results for TT are in Figure \ref{fig:TT_spectras_pred_vs_truth}, for EE in Figure \ref{fig:EE_spectras_pred_vs_truth}, for TE in Figure \ref{fig:TE_spectras_pred_vs_truth}, for PP in Figure~\ref{fig:PP_spectras_pred_vs_truth}, for PKL and PKNL in Figure \ref{fig:PKLNL_pred_vs_truth_ratio_in_percents}, for H, DA, and S8 in Figure \ref{fig:HDAs8_pred_vs_truth_ratio_in_percents}, and for the derived parameters in Figure \ref{fig:der_pred_vs_truth_ratio_in_percents}. For TT, TE, EE, and PP we show the difference in units of CMB-S4 sensitivity and in percentages, and for the other emulators we show the difference in percentages (since there is no simple way to compare with instrumental sensitivity; see comments in the previous paragraph). In all the figures, the three shades of red correspond to 68\%, 95\%, and 99\% of the testing data set, from dark to light, respectively. For instance, for TT (Figure \ref{fig:TT_spectras_pred_vs_truth}) we find that our emulators predict the power spectra to better than $0.02\sigma$ at CMB-S4 precision level, 68\% of the time in all the four cosmological models considered here. The agreement for TE and EE is similar. For PP the agreement is at the $0.02\sigma$ CMB-S4 precision level, 99\% of the time in all the four cosmological models considered here.

The quantity that is the least accurately emulated is the non-linear matter power spectrum, which is reproduced at the $1-1.5\%$ level 95\% of the time in all models (see Figure \ref{fig:PKLNL_pred_vs_truth_ratio_in_percents}). This level of agreement is satisfactory for application to current survey configurations, especially since non-linear modeling and baryonic uncertainty is at the $\approx 10\%$ level \citep[see, e.g.,][]{Mead_2021}  and even exceeds $30\%$ at $k>1\,\mathrm{Mpc}^{-1}$ \citep[see, e.g.,][]{Amon:2022azi}.

The emulated redshift-dependent $H(z)$, $D_A(z)$, and $\sigma_8(z)$ agree to better than 0.05\% between $z=0$ and $20$, for 99\% of the testing set in all models (see Figure \ref{fig:HDAs8_pred_vs_truth_ratio_in_percents}). We find a similar agreement between emulated derived parameters and the testing dataset (see Figure \ref{fig:der_pred_vs_truth_ratio_in_percents}).

\section{Accelerated likelihood analyses}\label{sec:fastan}

In this section we run  posterior inference analysis and use the emulators presented above in MCMC extractions of cosmological parameters. A full Stage-IV analysis will require optmization of likelihood codes alongside emulators which is beyond the scope of this paper and will be covered in future work. Therefore, here we demonstrate accelerated likelihood analyses for current (Stage II-III) data and reproduce marginalized constraints for ACT DR4 (Subsection \ref{ssec:actdr4}), \textit{Planck} lensing + DES + BAO (Subsection \ref{ssec:lensdesbao}) and \textit{Planck} TT,TE,EE in $w$CDM, $\Lambda$CDM+$N_\mathrm{eff}$ and   $\Lambda$CDM+$\Sigma m_\nu$ (Subsection \ref{ssec:exts}).
We use \textsc{cobaya} \citep{Torrado:2020dgo} for MCMC sampling. To analyse the resulting chains we use \verb|GetDist| \citep[][]{Lewis:2019xzd}.

We define a \textsc{CosmoPower} theory object within \textsc{cobaya} such that we are able to implement the \textsc{CosmoPower} call completely outside of the likelihoods. Our implementation of the \textsc{CosmoPower} theory object is available upon request. In a forthcoming paper, we plan to release an official \textsc{CosmoPower} wrapper for both \textsc{cobaya} and \textsc{cosmosis} \citep{Zuntz:2014csq}.

\subsection{Accelerated ACT DR4 power spectrum analysis}\label{ssec:actdr4}

We use the ACT Data Release 4 (DR4) ``actpol\_lite'' likelihood  \citep{ACT:2020frw,ACT:2020gnv} with \textsc{CosmoPower} to reproduce the original DR4 results.  This is the foreground-marginalized likelihood publicly available in the  \verb|pyactlike| repository\footnote{\href{https://github.com/ACTCollaboration/pyactlike}{https://github.com/ACTCollaboration/pyactlike}\label{fn:pyact}}. For the \textsc{CosmoPower} runs, this likelihood involves the TT, TE, and EE emulators, relying on multipoles up to $\ell\approx 4500$.

\begin{figure*}
    \includegraphics[width=2.\columnwidth]{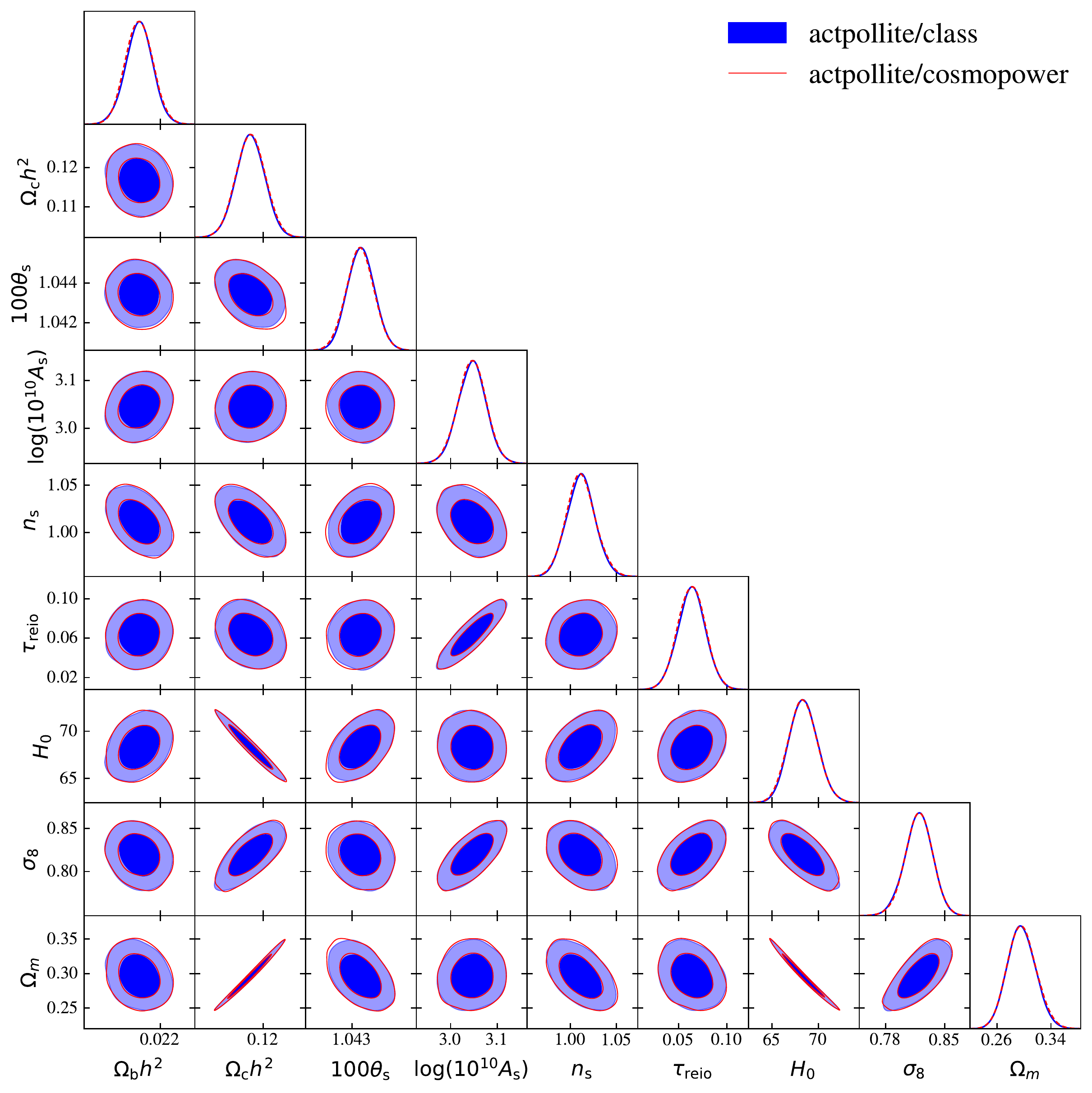}
    \vspace{0.2cm}
    \caption{Comparison of 2D marginalized posterior probability distributions for ACT DR4 cosmological parameters in $\Lambda$CDM, between \textsc{CosmoPower} (empty red contours) and the full \textsc{class} calculation (solid blue contours, taken from \citet{Hill:2021yec}). With \textsc{CosmoPower}, the derived parameters $H_0$, $\sigma_8$, and $\Omega_\mathrm{m}$ are added in post-processing of the chains. These results are obtained with the "actpol\_lite" ACT DR4 likelihood (see footnote \ref{fn:pyact}) and Subsection \ref{ssec:actdr4} for details.}
    \label{fig:contours_actdr4}
\end{figure*}

In Figure \ref{fig:contours_actdr4} we show the resulting 2D marginalized posterior probability distribution obtained in a few minutes on a laptop with \textsc{CosmoPower} as red empty contours, and compare it with the publicly available chains\footnote{CLASS2p8\_ACTPol\_lite\_DR4\_leakfix\_yp2\_baseLCDM\_taup\_hip\_R0p01 downloaded from \href{https://lambda.gsfc.nasa.gov/product/act/actpol_mcmc_chains_get.html}{LAMBDA}.}
shown as the filled green contours. The reference \textsc{class} chains were obtained in \cite{Hill:2021yec} using high-accuracy settings and took several days to converge. The overlap of the \textsc{CosmoPower} and reference contours is nearly perfect. We note that for the \textsc{CosmoPower} run we sample over $H_0$ and obtain $\theta_s$ as a derived parameter using our DER emulator in post-processing of the chains, whereas the reference \textsc{class} chains were computed with $\theta_s$ as an input parameter and $H_0$ was saved as a derived parameter.

\begin{figure*}
    \includegraphics[width=2.\columnwidth]{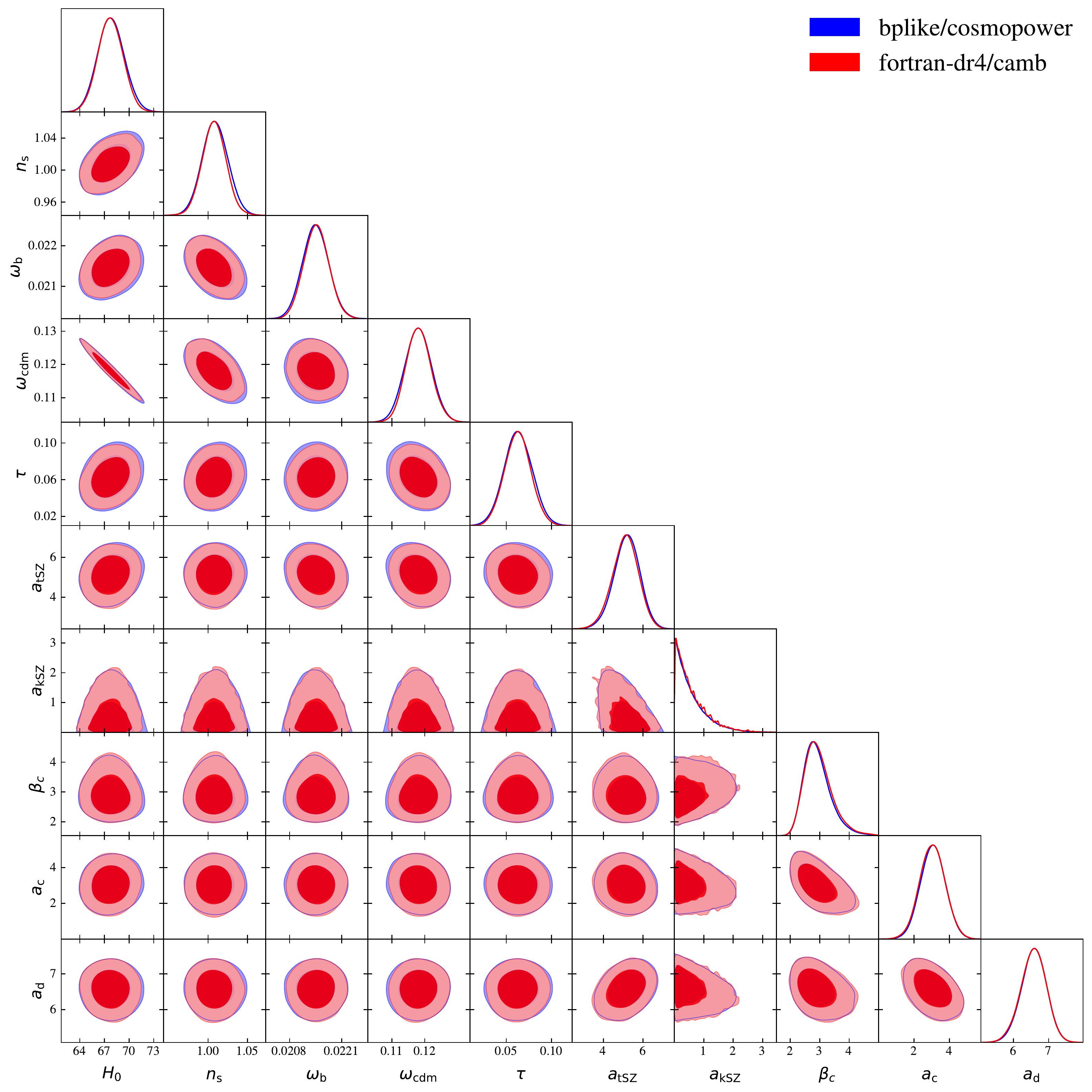}
    \vspace{0.2cm}
    \caption{Subset of cosmological and foreground parameters obtained with the ACT DR4 full likelihood (see \protect\cite{ACT:2020frw} for parameters definition). 2D marginalized posterior probability distributions from \textsc{CosmoPower} (blue contours) are compared to the ones obtained with the high-precision \textsc{camb} calculation (red contours).  See Subsection \ref{ssec:actdr4} for details.}
    \label{fig:contours_actdr4_fulllkl}
\end{figure*}

As a second test, we perform a maximum-likelihood analysis using the full ACT DR4 likelihood, including foregrounds. To do so we use a public python implementation \verb|bpllike|\footnote{\href{https://github.com/ACTCollaboration/bplike}{https://github.com/ACTCollaboration/bplike}}. The \textsc{CosmoPower} results (blue contours) are shown in Figure \ref{fig:contours_actdr4_fulllkl} and compared with the reference high-precision \textsc{camb} chains that we ran using the original full Fortran ACT DR4 likelihood\footnote{\href{https://lambda.gsfc.nasa.gov/product/act/act_dr4_likelihood_get.html}{ACT DR4 Fortran likelihood, named actpolfull there.}} (red contours).  Again, the overlap between the contours is nearly perfect. The foreground parameter contours (for the thermal SZ and kinematic SZ amplitudes and CIB SED power-law index and amplitude) match exactly. Cosmological parameter contours agree to better than  0.1$\sigma$, except for $\theta_s$. This is simply because of different definitions in \textsc{class} and \textsc{camb}. Note that our \textsc{cosmopower} chains reach $R-1=0.05$ in $\approx$20 minutes on a laptop with ten threads.

\subsection{Accelerated \textit{Planck} lensing + DES + BAO analysis}\label{ssec:lensdesbao}

To further demonstrate the efficiency and wide range of our emulators, we reproduce cosmological parameters extracted from the \emph{Planck} CMB lensing power spectrum + DES + BAO analysis.  We compare with the publicly available \emph{Planck} chains\footnote{\href{https://wiki.cosmos.esa.int/planck-legacy-archive/index.php/Cosmological_Parameters}{https://wiki.cosmos.esa.int/planck-legacy-archive}\label{fn:pla}}. The chains of interest are labeled  \texttt{DES\_lenspriors\_lensing\_BAO}.

We use the Python-native re-implementation of the \emph{Planck} reconstructed lensing power spectrum likelihood, assuming their fiducial (conservative) multipole range.  This likelihood is available in \textsc{cobaya} as \texttt{planck\_lensing\_2018} \citep{Planck:2018lbu}. We apply the priors used in the \emph{Planck} lensing analysis: Gaussian priors on $n_\mathrm{s}=0.96\pm0.02$ and $\Omega_\mathrm{b}h^2=0.0222\pm0.0005$, a flat prior $40 < H_0/({\rm km/s/Mpc}) < 100$, and a fixed $\tau=0.055$ \citep[see][for details]{Planck:2018lbu}.

We use the DES-Y1 cosmic shear + galaxy auto- and cross-correlation (``3$\times$2-point'') likelihood, as implemented in \texttt{des\_y1} in \textsc{cobaya}, which is as described by \cite{Troxel_2018}, \cite{Abbott_2018}, and \cite{DES:2017tss}.

Finally, we use the BAO likelihoods corresponding to BOSS DR12 \citep{Alam_2017}, the SDSS Main Galaxy Sample \citep{Ross:2014qpa}, and the 6dF survey \citep[][]{Beutler2011MNRAS.416.3017B}. These are implemented in \textsc{cobaya} as \texttt{sdss\_dr12\_consensus\_bao}, \texttt{sdss\_dr7\_mgs}, and \texttt{sixdf\_2011\_bao}.

For the \textsc{CosmoPower} runs, this likelihood combination involves the PKNL, H, DA, and PP emulators. Because this calculation requires the computation of the non-linear $P(k)$ at many redshifts, it is slightly more time-consuming. Indeed, we need to evaluate the matter power spectrum on a 2D grid in $(k,z)$, hence requiring us to call the emulator as many times as there are points in the $z$ dimension. We set the $P(k,z)$ grid to cover redshifts between 0 and 4 with 15 linearly spaced points. With this choice, \textsc{cobaya} performs $\approx$16 evaluations per second on an ARM64 MacBook Pro. We reach $R-1\simeq 0.15$ after $\approx$30 minutes on a laptop, with four chains taking a total of 56,000 accepted steps with acceptance rate of 0.2.

\begin{figure*}
    \includegraphics[width=2.\columnwidth]{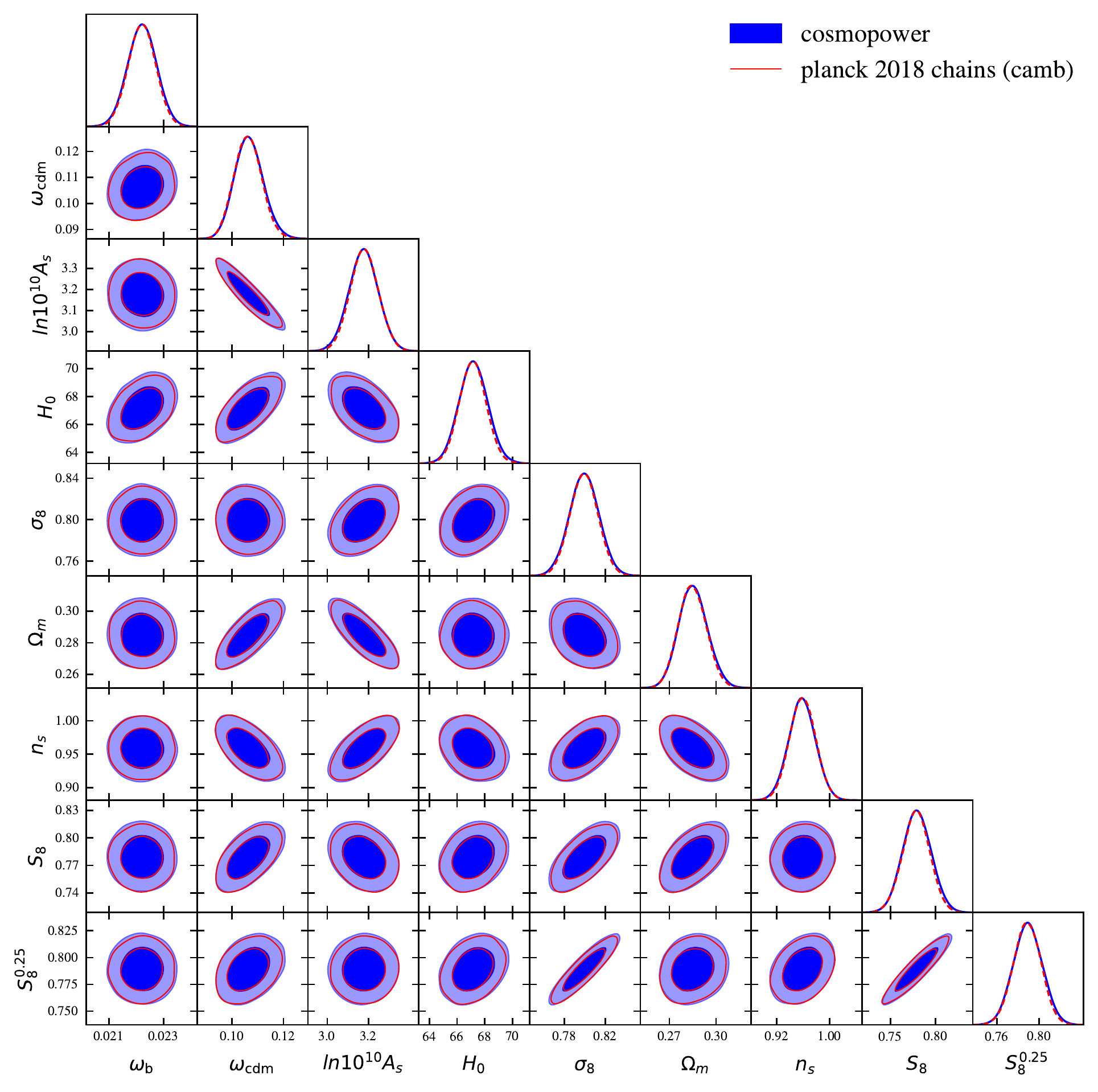}
    \vspace{0.2cm}
    \caption{ Comparison of 2D marginalized posterior probability distributions between \textsc{CosmoPower} (blue contours) and the reference  \protect\textsc{camb} chains (red contours, dashed line) downloaded from the \emph{Planck} legacy archive (footnote \ref{fn:pla}) for the \emph{Planck} 2018 CMB lensing likelihood  + DES-Y1 + BAO analysis. See Section \ref{ssec:lensdesbao} for details.
    }
    \label{fig:contours-des_lensing_bao}
\end{figure*}

The results are shown in Figure \ref{fig:contours-des_lensing_bao}. The overlap between the \textsc{CosmoPower} and reference \textsc{camb} contours is perfect. As derived parameters,  we show $S_8\equiv \sigma_8(\Omega_\mathrm{m}/0.3)^{0.5}$ and $S_8^{0.25}\equiv \sigma_8(\Omega_\mathrm{m}/0.3)^{0.25}$, which are most constrained by galaxy weak lensing and CMB lensing data, respectively. Our \textsc{CosmoPower} runs recover the reference constraints on  $S_8$ and $S_8^{0.25}$ exactly. This validates the matter power spectrum emulators for current galaxy WL surveys.

\subsection{RSD Full shape analysis}\label{ssec:fs8}

\begin{figure*}
    \includegraphics[width=2.\columnwidth]{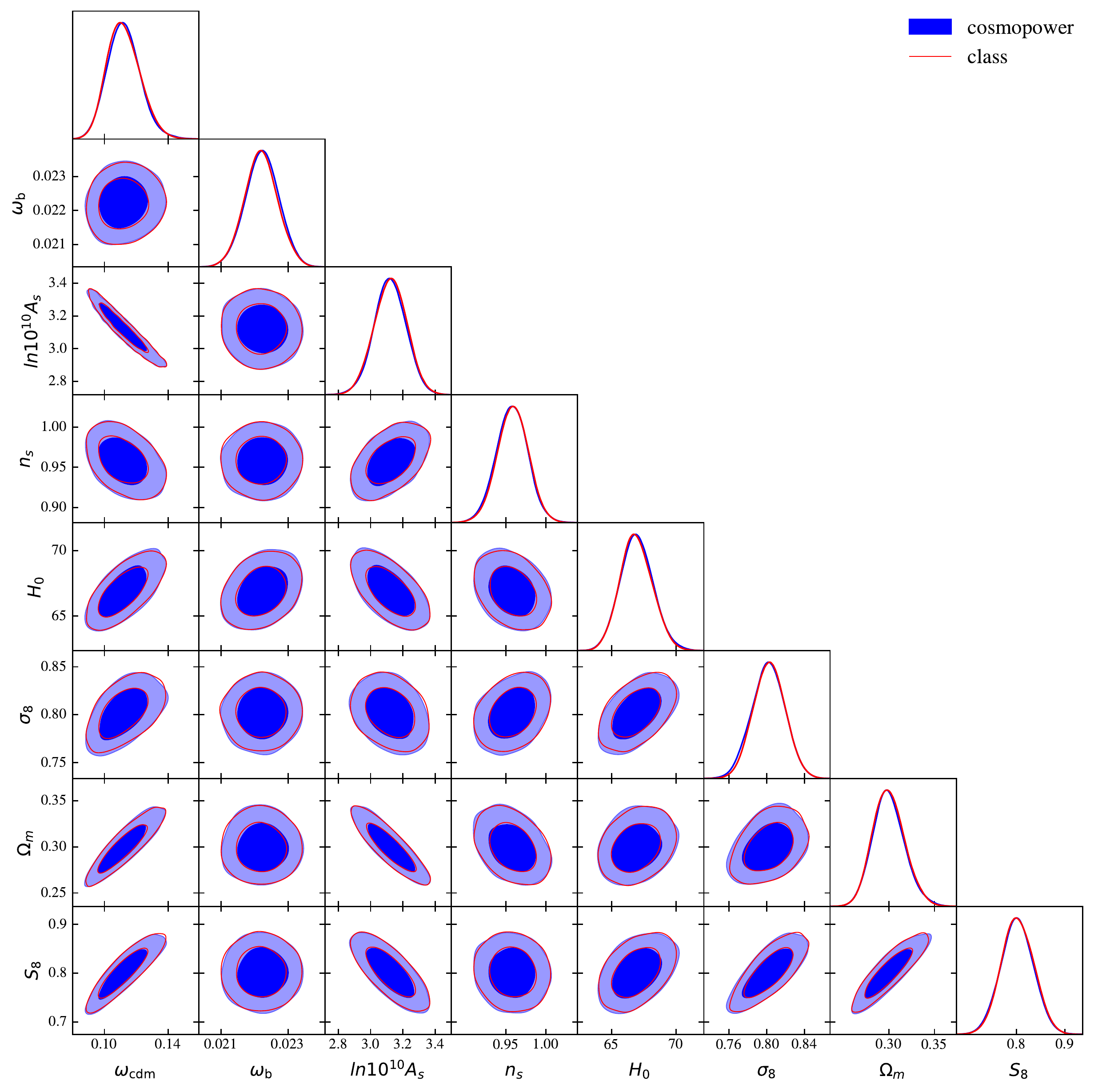}
    \vspace{0.2cm}
    \caption{ Comparison of 2D marginalized posterior probability distributions between \textsc{CosmoPower} (\emph{Planck} 2018 CMB lensing, the BOSS ``full shape'' likelihood including RSD information, to illustrate that $f\sigma_8$ can be computed with our emulators. See Section \ref{ssec:fs8} for details. This validates the S8 emulator.}
    \label{fig:contours-planck_lensfs8}
\end{figure*}

To demonstrate that we can also compute $f\sigma_8$ and use it in a likelihood analysis, using our S8 emulator (see Section \ref{sec:method}), we perform a likelihood analysis and use  \texttt{sdss\_dr12\_consensus\_fullshape} with \emph{Planck} lensing. We compare \textsc{class} and \textsc{CosmoPower} chains in Figure \ref{fig:contours-planck_lensfs8}.

\begin{figure*}
    \includegraphics[width=2.\columnwidth]{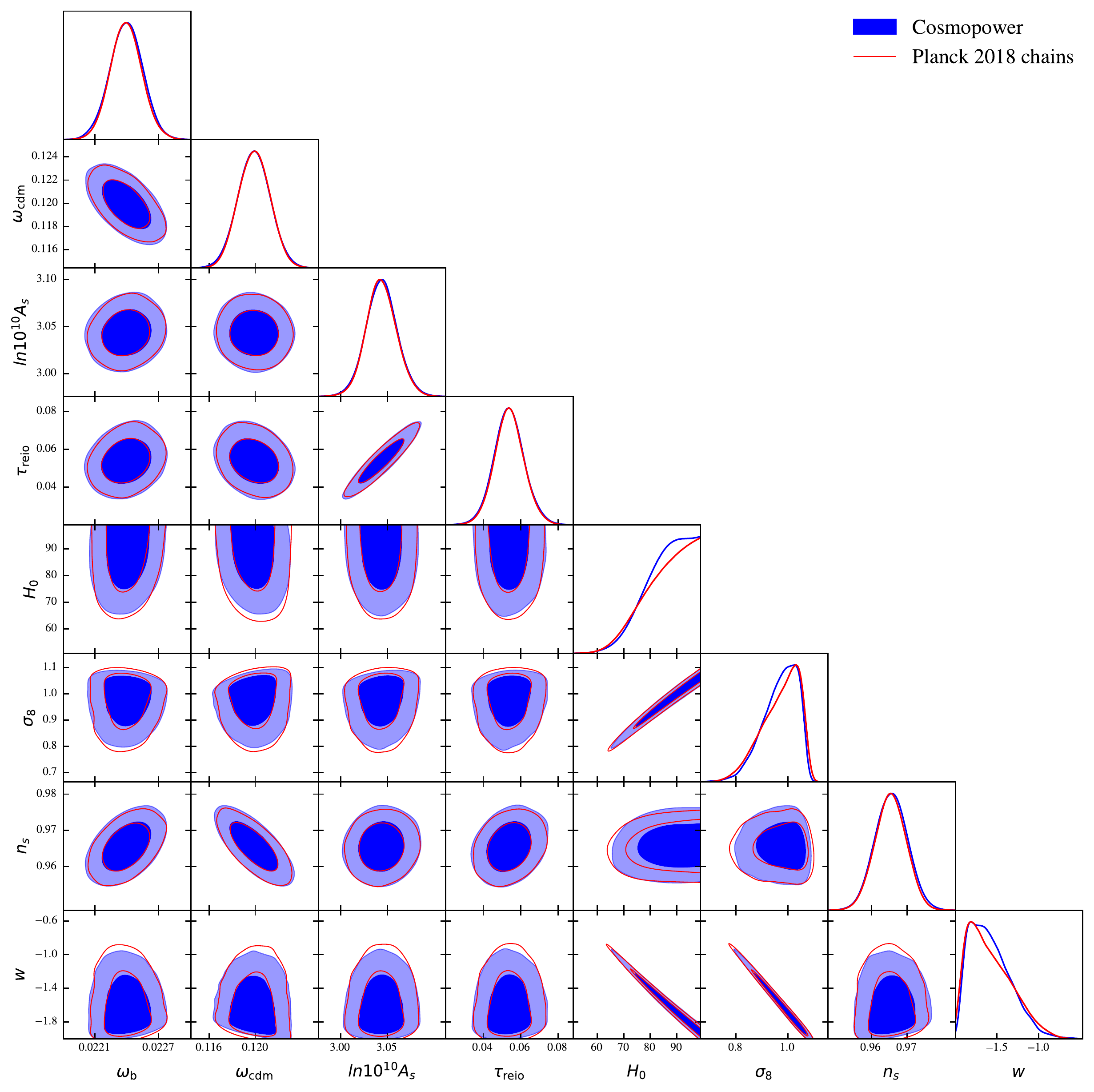}
    \vspace{0.2cm}
    \caption{ Comparison of 2D marginalized posterior probability distributions between \textsc{CosmoPower} (blue contours) and the full calculation (red contours) for the \emph{Planck} 2018 CMB likelihood \texttt{HM\_TTTEEE+lowl+lowE} computed using \protect\textsc{camb} in $w$CDM and downloaded from the \emph{Planck} legacy archive (footnote \ref{fn:pla}). See Subsection \ref{ssec:exts} for details. Note that our emulators cover the range $w>-2$ (which we used as a hard prior bound for our\textsc{CosmoPower} chains), as this is the lower bound adopted in the DES Year 1 analysis \citep{DES:2017tss}, although the \emph{Planck} chains allow for $w>-3$.}
    \label{fig:base_w}
\end{figure*}

\begin{figure*}
    \includegraphics[width=2.\columnwidth]{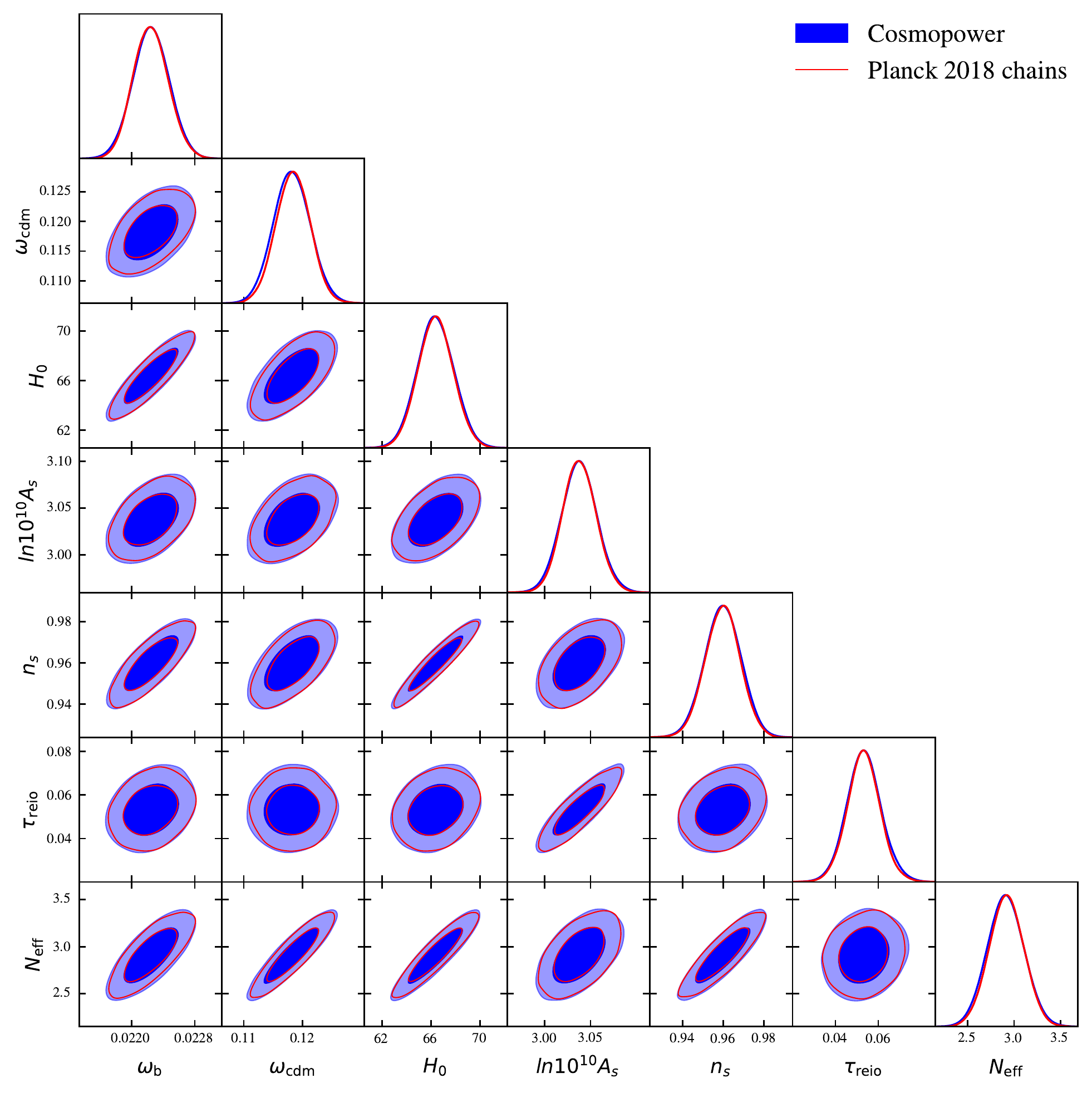}
    \vspace{0.2cm}
    \caption{ Comparison of 2D marginalized posterior probability distributions between \textsc{CosmoPower} (blue contours) and the full calculation (red contours) for the \emph{Planck} 2018 CMB likelihood \texttt{HM\_TTTEEE+lowl+lowE} (footnote \ref{fn:pla}) computed using \protect\textsc{camb} in $\Lambda$CDM+$N_\mathrm{eff}$ and downloaded from the \emph{Planck} legacy archive (footnote \ref{fn:pla}). See Subsection \ref{ssec:exts} for details.}
    \label{fig:base_nnu}
\end{figure*}

\begin{figure*}
    \includegraphics[width=2.\columnwidth]{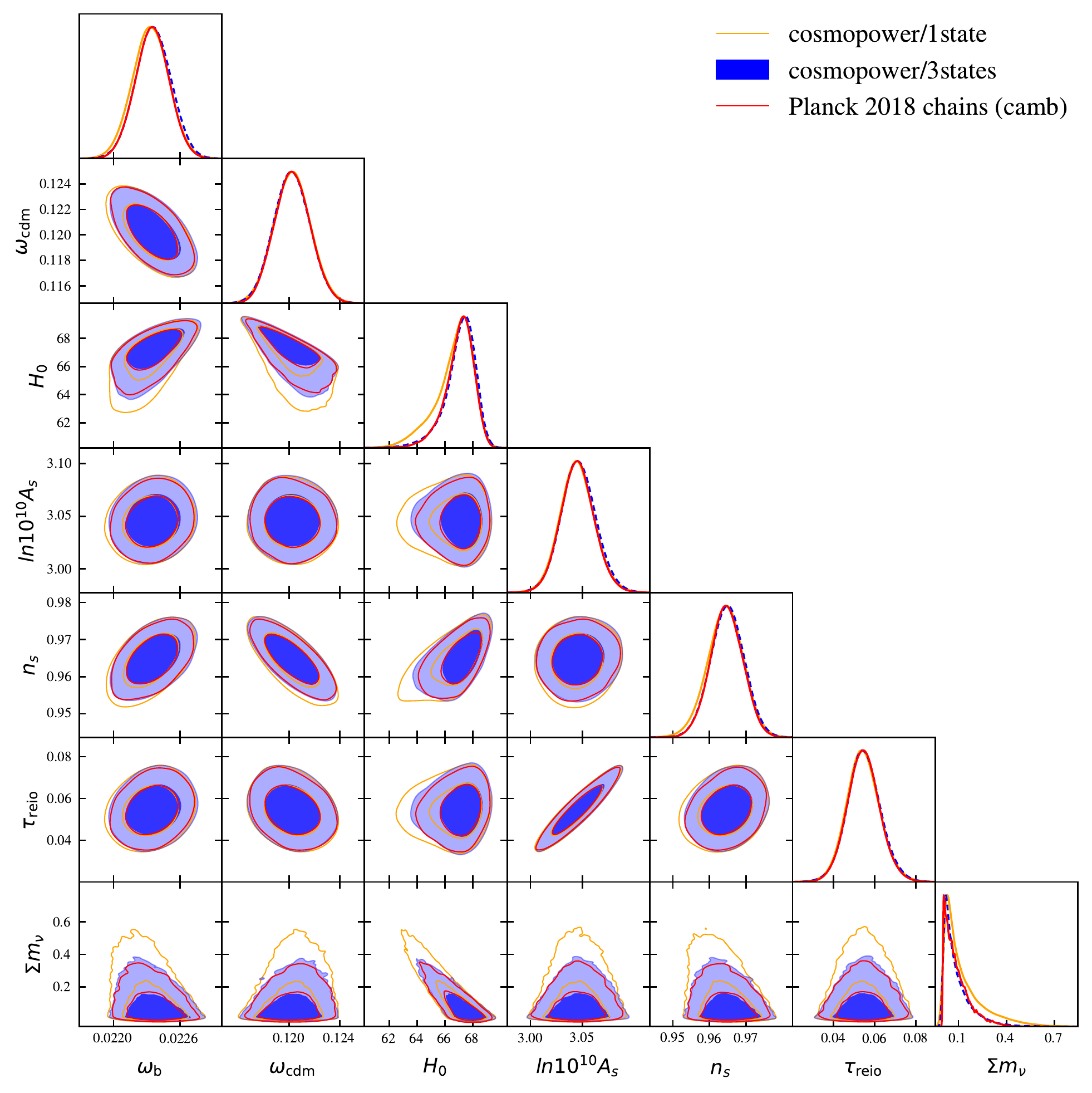}
    \vspace{0.2cm}
    \caption{Comparison of 2D marginalized posterior probability distributions between \textsc{CosmoPower} (blue contours) and the full calculation  for the \emph{Planck} 2018 CMB likelihood \texttt{HM\_TTTEEE+lowl+lowE}  computed using \protect\textsc{camb} (red contours) in $\Lambda$CDM+$\Sigma m_\nu$ and downloaded from the \emph{Planck} legacy archive (footnote \ref{fn:pla}). The yellow contours are the constraints using the $\Lambda$CDM+$\Sigma m_\nu$ emulator with one massive and two massless neutrinos (unlike the \textit{Planck} chains), rather than three massive neutrinos (blue and red contours). The model with three degenerate states is a better approximation to the mass splitting results. (We thank Antony Lewis for pointing this out to us.) See Subsection \ref{ssec:exts} for further details.}
    \label{fig:base_mnu}
\end{figure*}

\subsection{Extensions}\label{ssec:exts}

To validate our emulators in extended cosmology we run maximum-likelihood analyses in $w$CDM, $\Lambda$CDM+$N_\mathrm{eff}$, and $\Lambda$CDM+$\Sigma m_\nu$.
For the data and likelihood we choose the official \verb|clik| \emph{Planck} 2018 \texttt{plikHMTTTEEE+lowl+lowE} \citep{plc_2020} implementation, which we call through \textsc{cobaya}.

We show the  $w$CDM, $\Lambda$CDM+$N_\mathrm{eff}$, and $\Lambda$CDM+$\Sigma m_\nu$ constraints in Figure \ref{fig:base_w},  \ref{fig:base_nnu}, and \ref{fig:base_mnu}, respectively.  In each extension, we overplot the reference contours from \textsc{camb} chains (empty red contours) that we obtain online (see footnote \ref{fn:pla}). Our \textsc{CosmoPower} contours are in blue. In all cases the agreement with the reference \emph{Planck} chains is excellent. To reach $R-1\approx 0.1$, \textsc{CosmoPower} needs approximately 25 minutes, while \textsc{class} or \textsc{camb} chains typically take $\mathcal{O}(1\,\mathrm{day})$ to converge.

\section{Conclusions}\label{sec:end}

We have extended the work of \citet{Spurio_Mancini_2022} by creating emulators for CMB temperature and polarization angular anisotropy power spectra, CMB lensing convergence power spectra, linear and non-linear matter power spectra, and the redshift evolution of $H(z)$, $D_A(z)$ and $\sigma_8(z)$, in $\Lambda$CDM and extensions, namely, $w$CDM, $\Lambda$CDM+$N_\mathrm{eff}$ and  $\Lambda$CDM+$\Sigma m_\nu$ with one or three massive states (see Section \ref{sec:method}). All of these quantities are computed using high-precision \textsc{class} settings (see Section \ref{sec:emulacc}). They are sufficiently precise for Stage-IV CMB analyses (see Section \ref{sec:boltacc}) and current galaxy weak lensing and clustering surveys. We have tested all of our emulators in maximum-likelihood analyses, involving high-$\ell$ CMB, galaxy weak lensing and clustering, BAO and RSD data (see Section \ref{sec:fastan}).

The main outcome of this work is to open the door to fast parameter inference, via the widely used MCMC method, with accelerated Boltzmann computations thanks to our \textsc{CosmoPower} emulators. For instance, as we demonstrate here, most of Stage-III parameter inference is now feasible on a laptop.

In a forthcoming paper we plan release wrappers of our emulators so that they can be readily used within \textsc{cobaya} \citep{Torrado:2020dgo} and \textsc{cosmosis} \citep{Zuntz:2014csq}.

We created an online repository to store our emulators and will continue updating it. For CMB polarization, we have made emulators for E modes, which could be used to accelerate \textit{LiteBIRD}~\citep{2022arXiv220202773L} analyses dedicated on reionization history constraints \citep{Zaldarriaga:2008ap,LiteBIRD:2022cnt}. We defer B-modes power spectra emulators to future work.

In addition to the applications considered here, these emulators can also be used to save time at the initial step of libraries for computations of cosmological LSS observables like \textsc{class\_sz} \citep{Bolliet:2017lha,Bolliet:2022pze}, \textsc{ccl}  \citep{LSSTDarkEnergyScience:2018yem}, and any other code which relies on the quantities emulated here.

Recent works \citep{Philcox:2021kcw,Cabass:2022wjy} have used the effective field theory of large-scale structure \citep{Baumann:2010tm,Carrasco:2012cv} in order to derive constraints from spectroscopic surveys, based on the one-loop galaxy power spectrum and bispectrum. These calculations are time-consuming (MCMC convergence takes $\mathcal{O}$(1day) using \textsc{class\_pt} \citep[][]{Chudaykin:2020aoj}. The bottleneck is the calculation of higher-order correlators. The work presented here does not directly help to accelerate such analyses, but nonetheless the method can directly be applied in this context as well.

\section*{Acknowledgements}

 We thank Antony Lewis for many valuable discussions which improved the work presented here, as well as Ian Harrison for comments on the manuscript. We thank Nick Carriero and Robert Blackwell for key computational advice.

We acknowledge the use of computational resources at the Flatiron Institute.  The Flatiron Institute is supported by the Simons Foundation. ASM acknowledges support from the MSSL STFC Consolidated Grant ST/W001136/1. This work was partially enabled by funding from the University College London (UCL) Cosmoparticle Initiative and by collaborative visits funded by Princeton University and the Cosmology and Astroparticle Student and Postdoc Exchange Network (CASPEN). ASM thanks Princeton University and the Flatiron Institute for their hospitality during his visit, during which this project was started. JCH acknowledges support from NSF grant AST-2108536, NASA grant 21-ATP21-0129, DOE grant DE-SC00233966, the Sloan Foundation, and the Simons Foundation.  BB acknowledges  support from the European Research Council (ERC) under the European Union’s Horizon 2020 research and innovation programme (Grant agreement No. 851274). EC and HJ acknowledge support from the European Research Council (ERC) under the European Union’s Horizon 2020 research and innovation programme (Grant agreement No. 849169). JD acknowledges support from NSF grant AST-2108126.

\section*{Author contributions}

BB led the work. ASM contributed to the development of the project and software pipeline, helped define the library of models and parameters to be studied, and contributed to the writing and editing of the manuscript. JCH contributed to studies of numerical precision settings, comparisons of \textsc{camb} and \textsc{class}, contributed to the writing and editing of the manuscript, and coordinated the initiation and overall direction of this work. MM aided in investigations into precision settings and prior ranges required for various parameters, helped define the library of models to be produced, developed the \texttt{bplike} software for the ACT DR4 likelihood analysis and contributed to the writing and editing of the manuscript. HTJ is developing the wrappers that will import these emulators into future ACT and SO analyses, and reviewed the paper at the final stages and provided feedback. EC developed the ACT DR4 likelihood, contributed to discussions and to the writing and editing of the manuscript. JD coordinated the initiation of this work, and contributed to the editing of the manuscript.

\section*{Data Availability}

The emulators are available online at \href{https://github.com/cosmopower-organization}{CosmoPower Organisation}, including tutorial notebooks. If you use the emulators, please cite our work as well as \cite{Spurio_Mancini_2022}.



\bibliographystyle{mnras}
\bibliography{mnras_template} 

\begin{thebibliography}{}
\makeatletter
\relax
\def\mn@urlcharsother{\let\do\@makeother \do\$\do\&\do\#\do\^\do\_\do\%\do\~}
\def\mn@doi{\begingroup\mn@urlcharsother \@ifnextchar [ {\mn@doi@}
  {\mn@doi@[]}}
\def\mn@doi@[#1]#2{\def\@tempa{#1}\ifx\@tempa\@empty \href
  {http://dx.doi.org/#2} {doi:#2}\else \href {http://dx.doi.org/#2} {#1}\fi
  \endgroup}
\def\mn@eprint#1#2{\mn@eprint@#1:#2::\@nil}
\def\mn@eprint@arXiv#1{\href {http://arxiv.org/abs/#1} {{\tt arXiv:#1}}}
\def\mn@eprint@dblp#1{\href {http://dblp.uni-trier.de/rec/bibtex/#1.xml}
  {dblp:#1}}
\def\mn@eprint@#1:#2:#3:#4\@nil{\def\@tempa {#1}\def\@tempb {#2}\def\@tempc
  {#3}\ifx \@tempc \@empty \let \@tempc \@tempb \let \@tempb \@tempa \fi \ifx
  \@tempb \@empty \def\@tempb {arXiv}\fi \@ifundefined
  {mn@eprint@\@tempb}{\@tempb:\@tempc}{\expandafter \expandafter \csname
  mn@eprint@\@tempb\endcsname \expandafter{\@tempc}}}

\bibitem[\protect\citeauthoryear{Abadi et~al.,}{Abadi
  et~al.}{2015}]{tensorflow2015-whitepaper}
Abadi M.,  et~al., 2015, {TensorFlow}: Large-Scale Machine Learning on
  Heterogeneous Systems, \url {https://www.tensorflow.org/}

\bibitem[\protect\citeauthoryear{{Abazajian} et~al.,}{{Abazajian}
  et~al.}{2019}]{CMBS4_DSR}
{Abazajian} K.,  et~al., 2019, \mn@doi [arXiv e-prints]
  {10.48550/arXiv.1907.04473}, \href
  {https://ui.adsabs.harvard.edu/abs/2019arXiv190704473A} {p. arXiv:1907.04473}

\bibitem[\protect\citeauthoryear{{Aihara} et~al.,}{{Aihara}
  et~al.}{2018}]{2018PASJ...70S...4A}
{Aihara} H.,  et~al., 2018, \mn@doi [\pasj] {10.1093/pasj/psx066}, \href
  {https://ui.adsabs.harvard.edu/abs/2018PASJ...70S...4A} {70, S4}

\bibitem[\protect\citeauthoryear{Aiola et~al.}{Aiola
  et~al.}{2020}]{ACT:2020gnv}
Aiola S.,  et~al., 2020, \mn@doi [JCAP] {10.1088/1475-7516/2020/12/047}, 12,
  047

\bibitem[\protect\citeauthoryear{{Akeson} et~al.,}{{Akeson}
  et~al.}{2019}]{2019arXiv190205569A}
{Akeson} R.,  et~al., 2019, \mn@doi [arXiv e-prints]
  {10.48550/arXiv.1902.05569}, \href
  {https://ui.adsabs.harvard.edu/abs/2019arXiv190205569A} {p. arXiv:1902.05569}

\bibitem[\protect\citeauthoryear{Alam et~al.}{Alam et~al.}{2017}]{Alam_2017}
Alam S.,  et~al., 2017, \mn@doi [Mon. Not. Roy. Astron. Soc.]
  {10.1093/mnras/stx721}, 470, 2617

\bibitem[\protect\citeauthoryear{Albers, Fidler, Lesgourgues, Sch\"oneberg  \&
  Torrado}{Albers et~al.}{2019}]{Albers:2019rzt}
Albers J.,  Fidler C.,  Lesgourgues J.,  Sch\"oneberg N.,   Torrado J.,  2019,
  \mn@doi [JCAP] {10.1088/1475-7516/2019/09/028}, 09, 028

\bibitem[\protect\citeauthoryear{Allys et~al.}{Allys
  et~al.}{2022}]{LiteBIRD:2022cnt}
Allys E.,  et~al., 2022, \mn@doi [] {10.1093/ptep/ptac150}

\bibitem[\protect\citeauthoryear{{Amon} \& {Efstathiou}}{{Amon} \&
  {Efstathiou}}{2022}]{Amon:2022azi}
{Amon} A.,  {Efstathiou} G.,  2022, \mn@doi [\mnras] {10.1093/mnras/stac2429},
  \href {https://ui.adsabs.harvard.edu/abs/2022MNRAS.516.5355A} {516, 5355}

\bibitem[\protect\citeauthoryear{{Aric{\`o}}, {Angulo}  \&
  {Zennaro}}{{Aric{\`o}} et~al.}{2021}]{Arico21}
{Aric{\`o}} G.,  {Angulo} R.~E.,   {Zennaro} M.,  2021, arXiv e-prints, \href
  {https://ui.adsabs.harvard.edu/abs/2021arXiv210414568A} {p. arXiv:2104.14568}

\bibitem[\protect\citeauthoryear{{Auld}, {Bridges}, {Hobson}  \& {Gull}}{{Auld}
  et~al.}{2007}]{Auld07}
{Auld} T.,  {Bridges} M.,  {Hobson} M.~P.,   {Gull} S.~F.,  2007, \mn@doi
  [\mnras] {10.1111/j.1745-3933.2006.00276.x}, \href
  {https://ui.adsabs.harvard.edu/abs/2007MNRAS.376L..11A} {376, L11}

\bibitem[\protect\citeauthoryear{{Balkenhol} et~al.,}{{Balkenhol}
  et~al.}{2022}]{SPT-3G:2022hvq}
{Balkenhol} L.,  et~al., 2022, \mn@doi [arXiv e-prints]
  {10.48550/arXiv.2212.05642}, \href
  {https://ui.adsabs.harvard.edu/abs/2022arXiv221205642B} {p. arXiv:2212.05642}

\bibitem[\protect\citeauthoryear{Baumann, Nicolis, Senatore  \&
  Zaldarriaga}{Baumann et~al.}{2012}]{Baumann:2010tm}
Baumann D.,  Nicolis A.,  Senatore L.,   Zaldarriaga M.,  2012, \mn@doi [JCAP]
  {10.1088/1475-7516/2012/07/051}, 07, 051

\bibitem[\protect\citeauthoryear{{Benson} et~al.,}{{Benson}
  et~al.}{2014}]{2014SPIE.9153E..1PB}
{Benson} B.~A.,  et~al., 2014, in {Holland} W.~S.,  {Zmuidzinas} J.,  eds,
  Society of Photo-Optical Instrumentation Engineers (SPIE) Conference Series
  Vol. 9153, Millimeter, Submillimeter, and Far-Infrared Detectors and
  Instrumentation for Astronomy VII. p. 91531P (\mn@eprint {arXiv}
  {1407.2973}), \mn@doi{10.1117/12.2057305}

\bibitem[\protect\citeauthoryear{{Beutler} et~al.,}{{Beutler}
  et~al.}{2011}]{Beutler2011MNRAS.416.3017B}
{Beutler} F.,  et~al., 2011, \mn@doi [\mnras]
  {10.1111/j.1365-2966.2011.19250.x}, \href
  {https://ui.adsabs.harvard.edu/abs/2011MNRAS.416.3017B} {416, 3017}

\bibitem[\protect\citeauthoryear{{Blas}, {Lesgourgues}  \& {Tram}}{{Blas}
  et~al.}{2011}]{classII}
{Blas} D.,  {Lesgourgues} J.,   {Tram} T.,  2011, \mn@doi [\jcap]
  {10.1088/1475-7516/2011/07/034}, \href
  {https://ui.adsabs.harvard.edu/abs/2011JCAP...07..034B} {2011, 034}

\bibitem[\protect\citeauthoryear{Bolliet, Comis, Komatsu  \&
  Mac\'\i{}as-P\'erez}{Bolliet et~al.}{2018}]{Bolliet:2017lha}
Bolliet B.,  Comis B.,  Komatsu E.,   Mac\'\i{}as-P\'erez J.~F.,  2018, \mn@doi
  [Mon. Not. Roy. Astron. Soc.] {10.1093/mnras/sty823}, 477, 4957

\bibitem[\protect\citeauthoryear{{Bolliet}, {Hill}, {Ferraro}, {Kusiak}  \&
  {Krolewski}}{{Bolliet} et~al.}{2022}]{Bolliet:2022pze}
{Bolliet} B.,  {Hill} J.~C.,  {Ferraro} S.,  {Kusiak} A.,   {Krolewski} A.,
  2022, \mn@doi [arXiv e-prints] {10.48550/arXiv.2208.07847}, \href
  {https://ui.adsabs.harvard.edu/abs/2022arXiv220807847B} {p. arXiv:2208.07847}

\bibitem[\protect\citeauthoryear{CMBS4}{CMBS4}{2022}]{Abazajian_2022}
CMBS4 2022, \mn@doi [The Astrophysical Journal] {10.3847/1538-4357/ac1596},
  926, 54

\bibitem[\protect\citeauthoryear{Cabass, Ivanov, Philcox, Simonovi\'c  \&
  Zaldarriaga}{Cabass et~al.}{2022}]{Cabass:2022wjy}
Cabass G.,  Ivanov M.~M.,  Philcox O. H.~E.,  Simonovi\'c M.,   Zaldarriaga M.,
   2022, \mn@doi [Phys. Rev. Lett.] {10.1103/PhysRevLett.129.021301}, 129,
  021301

\bibitem[\protect\citeauthoryear{Carrasco, Hertzberg  \& Senatore}{Carrasco
  et~al.}{2012}]{Carrasco:2012cv}
Carrasco J. J.~M.,  Hertzberg M.~P.,   Senatore L.,  2012, \mn@doi [JHEP]
  {10.1007/JHEP09(2012)082}, 09, 082

\bibitem[\protect\citeauthoryear{Chisari et~al.}{Chisari
  et~al.}{2019}]{LSSTDarkEnergyScience:2018yem}
Chisari N.~E.,  et~al., 2019, \mn@doi [Astrophys. J. Suppl.]
  {10.3847/1538-4365/ab1658}, 242, 2

\bibitem[\protect\citeauthoryear{Choi et~al.}{Choi et~al.}{2020}]{ACT:2020frw}
Choi S.~K.,  et~al., 2020, \mn@doi [JCAP] {10.1088/1475-7516/2020/12/045}, 12,
  045

\bibitem[\protect\citeauthoryear{Chudaykin, Ivanov, Philcox  \&
  Simonovi\'c}{Chudaykin et~al.}{2020}]{Chudaykin:2020aoj}
Chudaykin A.,  Ivanov M.~M.,  Philcox O. H.~E.,   Simonovi\'c M.,  2020,
  \mn@doi [Phys. Rev. D] {10.1103/PhysRevD.102.063533}, 102, 063533

\bibitem[\protect\citeauthoryear{{DES Collaboration}}{{DES
  Collaboration}}{2018}]{Abbott_2018}
{DES Collaboration} 2018, \mn@doi [Physical Review D]
  {10.1103/physrevd.98.043526}, 98

\bibitem[\protect\citeauthoryear{{DESI Collaboration} et~al.,}{{DESI
  Collaboration} et~al.}{2016}]{DESI:2016fyo}
{DESI Collaboration} et~al., 2016, \mn@doi [arXiv e-prints]
  {10.48550/arXiv.1611.00036}, \href
  {https://ui.adsabs.harvard.edu/abs/2016arXiv161100036D} {p. arXiv:1611.00036}

\bibitem[\protect\citeauthoryear{{Dark Energy Survey Collaboration}}{{Dark
  Energy Survey Collaboration}}{2016}]{2016MNRAS.460.1270D}
{Dark Energy Survey Collaboration} 2016, \mn@doi [\mnras]
  {10.1093/mnras/stw641}, \href
  {https://ui.adsabs.harvard.edu/abs/2016MNRAS.460.1270D} {460, 1270}

\bibitem[\protect\citeauthoryear{{Dawson} et~al.,}{{Dawson}
  et~al.}{2013}]{2013AJ....145...10D}
{Dawson} K.~S.,  et~al., 2013, \mn@doi [\aj] {10.1088/0004-6256/145/1/10},
  \href {https://ui.adsabs.harvard.edu/abs/2013AJ....145...10D} {145, 10}

\bibitem[\protect\citeauthoryear{{Dor{\'e}} et~al.,}{{Dor{\'e}}
  et~al.}{2014}]{Dore:2014cca}
{Dor{\'e}} O.,  et~al., 2014, \mn@doi [arXiv e-prints]
  {10.48550/arXiv.1412.4872}, \href
  {https://ui.adsabs.harvard.edu/abs/2014arXiv1412.4872D} {p. arXiv:1412.4872}

\bibitem[\protect\citeauthoryear{Fendt \& Wandelt}{Fendt \&
  Wandelt}{2007}]{Fendt07}
Fendt W.~A.,  Wandelt B.~D.,  2007, \mn@doi [The Astrophysical Journal]
  {10.1086/508342}, 654, 2–11

\bibitem[\protect\citeauthoryear{G\"unther, Lesgourgues, Samaras, Sch\"oneberg,
  Stadtmann, Fidler  \& Torrado}{G\"unther et~al.}{2022}]{Gunther:2022pto}
G\"unther S.,  Lesgourgues J.,  Samaras G.,  Sch\"oneberg N.,  Stadtmann F.,
  Fidler C.,   Torrado J.,  2022, \mn@doi [JCAP]
  {10.1088/1475-7516/2022/11/035}, 11, 035

\bibitem[\protect\citeauthoryear{Henderson et~al.}{Henderson
  et~al.}{2016}]{Henderson:2015nzj}
Henderson S.~W.,  et~al., 2016, \mn@doi [J. Low Temp. Phys.]
  {10.1007/s10909-016-1575-z}, 184, 772

\bibitem[\protect\citeauthoryear{Hill et~al.}{Hill et~al.}{2022}]{Hill:2021yec}
Hill J.~C.,  et~al., 2022, \mn@doi [Phys. Rev. D]
  {10.1103/PhysRevD.105.123536}, 105, 123536

\bibitem[\protect\citeauthoryear{Ivezi\'c et~al.}{Ivezi\'c
  et~al.}{2019}]{LSST:2008ijt}
Ivezi\'c v.,  et~al., 2019, \mn@doi [Astrophys. J.] {10.3847/1538-4357/ab042c},
  873, 111

\bibitem[\protect\citeauthoryear{{Krause} et~al.,}{{Krause}
  et~al.}{2017}]{DES:2017tss}
{Krause} E.,  et~al., 2017, \mn@doi [arXiv e-prints]
  {10.48550/arXiv.1706.09359}, \href
  {https://ui.adsabs.harvard.edu/abs/2017arXiv170609359K} {p. arXiv:1706.09359}

\bibitem[\protect\citeauthoryear{{Kuijken} et~al.,}{{Kuijken}
  et~al.}{2015}]{2015MNRAS.454.3500K}
{Kuijken} K.,  et~al., 2015, \mn@doi [\mnras] {10.1093/mnras/stv2140}, \href
  {https://ui.adsabs.harvard.edu/abs/2015MNRAS.454.3500K} {454, 3500}

\bibitem[\protect\citeauthoryear{{Laureijs} et~al.,}{{Laureijs}
  et~al.}{2011}]{EUCLID:2011zbd}
{Laureijs} R.,  et~al., 2011, \mn@doi [arXiv e-prints]
  {10.48550/arXiv.1110.3193}, \href
  {https://ui.adsabs.harvard.edu/abs/2011arXiv1110.3193L} {p. arXiv:1110.3193}

\bibitem[\protect\citeauthoryear{{Lesgourgues}}{{Lesgourgues}}{2011a}]{classI}
{Lesgourgues} J.,  2011a, arXiv e-prints, \href
  {https://ui.adsabs.harvard.edu/abs/2011arXiv1104.2932L} {p. arXiv:1104.2932}

\bibitem[\protect\citeauthoryear{{Lesgourgues}}{{Lesgourgues}}{2011b}]{Lesgourgues_2011_CLASSIII}
{Lesgourgues} J.,  2011b, \mn@doi [arXiv e-prints] {10.48550/arXiv.1104.2934},
  \href {https://ui.adsabs.harvard.edu/abs/2011arXiv1104.2934L} {p.
  arXiv:1104.2934}

\bibitem[\protect\citeauthoryear{{Lewis}}{{Lewis}}{2019}]{Lewis:2019xzd}
{Lewis} A.,  2019, \mn@doi [arXiv e-prints] {10.48550/arXiv.1910.13970}, \href
  {https://ui.adsabs.harvard.edu/abs/2019arXiv191013970L} {p. arXiv:1910.13970}

\bibitem[\protect\citeauthoryear{{Lewis}, {Challinor}  \& {Lasenby}}{{Lewis}
  et~al.}{2000}]{camb2000ApJ...538..473L}
{Lewis} A.,  {Challinor} A.,   {Lasenby} A.,  2000, \mn@doi [\apj]
  {10.1086/309179}, \href
  {https://ui.adsabs.harvard.edu/abs/2000ApJ...538..473L} {538, 473}

\bibitem[\protect\citeauthoryear{{LiteBIRD Collaboration}}{{LiteBIRD
  Collaboration}}{2022}]{2022arXiv220202773L}
{LiteBIRD Collaboration} 2022, \mn@doi [arXiv e-prints]
  {10.48550/arXiv.2202.02773}, \href
  {https://ui.adsabs.harvard.edu/abs/2022arXiv220202773L} {p. arXiv:2202.02773}

\bibitem[\protect\citeauthoryear{McCarthy, Hill  \& Madhavacheril}{McCarthy
  et~al.}{2022}]{McCarthy:2021lfp}
McCarthy F.,  Hill J.~C.,   Madhavacheril M.~S.,  2022, \mn@doi [Phys. Rev. D]
  {10.1103/PhysRevD.105.023517}, 105, 023517

\bibitem[\protect\citeauthoryear{Mead, Peacock, Heymans, Joudaki  \&
  Heavens}{Mead et~al.}{2015}]{Mead15}
Mead A.~J.,  Peacock J.~A.,  Heymans C.,  Joudaki S.,   Heavens A.~F.,  2015,
  \mn@doi [Monthly Notices of the Royal Astronomical Society]
  {10.1093/mnras/stv2036}, 454, 1958

\bibitem[\protect\citeauthoryear{Mead, Brieden, Tröster  \& Heymans}{Mead
  et~al.}{2021}]{Mead_2021}
Mead A.~J.,  Brieden S.,  Tröster T.,   Heymans C.,  2021, \mn@doi [Monthly
  Notices of the Royal Astronomical Society] {10.1093/mnras/stab082}, 502, 1401

\bibitem[\protect\citeauthoryear{Mootoovaloo, Jaffe, Heavens  \&
  Leclercq}{Mootoovaloo et~al.}{2022}]{Mootoovaloo21}
Mootoovaloo A.,  Jaffe A.,  Heavens A.,   Leclercq F.,  2022, \mn@doi
  [Astronomy and Computing] {https://doi.org/10.1016/j.ascom.2021.100508}, 38,
  100508

\bibitem[\protect\citeauthoryear{Philcox \& Ivanov}{Philcox \&
  Ivanov}{2022}]{Philcox:2021kcw}
Philcox O. H.~E.,  Ivanov M.~M.,  2022, \mn@doi [Phys. Rev. D]
  {10.1103/PhysRevD.105.043517}, 105, 043517

\bibitem[\protect\citeauthoryear{{Planck Collaboration}}{{Planck
  Collaboration}}{2020a}]{plc_2020}
{Planck Collaboration} 2020a, \mn@doi [\aap] {10.1051/0004-6361/201833910},
  \href {https://ui.adsabs.harvard.edu/abs/2020A&A...641A...6P} {641, A6}

\bibitem[\protect\citeauthoryear{{Planck Collaboration}}{{Planck
  Collaboration}}{2020b}]{Planck:2018lbu}
{Planck Collaboration} 2020b, \mn@doi [Astron. Astrophys.]
  {10.1051/0004-6361/201833886}, 641, A8

\bibitem[\protect\citeauthoryear{Ross, Samushia, Howlett, Percival, Burden  \&
  Manera}{Ross et~al.}{2015}]{Ross:2014qpa}
Ross A.~J.,  Samushia L.,  Howlett C.,  Percival W.~J.,  Burden A.,   Manera
  M.,  2015, \mn@doi [Mon. Not. Roy. Astron. Soc.] {10.1093/mnras/stv154}, 449,
  835

\bibitem[\protect\citeauthoryear{{Sehgal} et~al.,}{{Sehgal} et~al.}{2019}]{hd}
{Sehgal} N.,  et~al., 2019, in Bulletin of the American Astronomical Society.
  p.~6 (\mn@eprint {arXiv} {1906.10134}), \mn@doi{10.48550/arXiv.1906.10134}

\bibitem[\protect\citeauthoryear{{Simons Observatory}}{{Simons
  Observatory}}{2019}]{Ade_2019}
{Simons Observatory} 2019, \mn@doi [Journal of Cosmology and Astroparticle
  Physics] {10.1088/1475-7516/2019/02/056}, 2019, 056

\bibitem[\protect\citeauthoryear{Smith et~al.,}{Smith
  et~al.}{2003}]{Smith_2003}
Smith R.~E.,  et~al., 2003, \mn@doi [Monthly Notices of the Royal Astronomical
  Society] {10.1046/j.1365-8711.2003.06503.x}, 341, 1311

\bibitem[\protect\citeauthoryear{{Spergel} et~al.,}{{Spergel}
  et~al.}{2015}]{2015arXiv150303757S}
{Spergel} D.,  et~al., 2015, \mn@doi [arXiv e-prints]
  {10.48550/arXiv.1503.03757}, \href
  {https://ui.adsabs.harvard.edu/abs/2015arXiv150303757S} {p. arXiv:1503.03757}

\bibitem[\protect\citeauthoryear{{Spurio~Mancini}, {Piras}, {Alsing},
  {Joachimi}  \& {Hobson}}{{Spurio~Mancini} et~al.}{2022}]{Spurio_Mancini_2022}
{Spurio~Mancini} A.,  {Piras} D.,  {Alsing} J.,  {Joachimi} B.,   {Hobson}
  M.~P.,  2022, \mn@doi [Monthly Notices of the Royal Astronomical Society]
  {10.1093/mnras/stac064}, 511, 1771

\bibitem[\protect\citeauthoryear{Torrado \& Lewis}{Torrado \&
  Lewis}{2021}]{Torrado:2020dgo}
Torrado J.,  Lewis A.,  2021, \mn@doi [JCAP] {10.1088/1475-7516/2021/05/057},
  05, 057

\bibitem[\protect\citeauthoryear{{Troxel} et~al.,}{{Troxel}
  et~al.}{2018}]{Troxel_2018}
{Troxel} M.~A.,  et~al., 2018, \mn@doi [\prd] {10.1103/PhysRevD.98.043528},
  \href {https://ui.adsabs.harvard.edu/abs/2018PhRvD..98d3528T} {98, 043528}

\bibitem[\protect\citeauthoryear{Zaldarriaga et~al.,}{Zaldarriaga
  et~al.}{2008}]{Zaldarriaga:2008ap}
Zaldarriaga M.,  et~al., 2008

\bibitem[\protect\citeauthoryear{Zuntz et~al.,}{Zuntz
  et~al.}{2015}]{Zuntz:2014csq}
Zuntz J.,  et~al., 2015, \mn@doi [Astron. Comput.]
  {10.1016/j.ascom.2015.05.005}, 12, 45

\makeatother
\end{thebibliography}

%


\bsp	
\label{lastpage}
\end{document}